\documentclass[preprint,prc,showpacs,preprintnumbers,amsmath,amssymb]{revtex4-1}
\usepackage{graphicx,latexsym,amssymb,amsmath,color,multirow,mathrsfs,booktabs,ifpdf}
\usepackage{dcolumn}
\usepackage{bm}

\usepackage{float, varioref}

\begin{document}
\title{Spin-polarized $\beta$-stable neutron star matter: The nuclear symmetry 
 energy and GW170817 constraint}
\author{Ngo Hai Tan$^{1,2}$}\email{tan.ngohai@phenikaa-uni.edu.vn}
\author{Dao T. Khoa$^3$}
\author{Doan Thi Loan$^3$}
\affiliation{$^1$ Faculty of Fundamental Sciences, Phenikaa University, Hanoi 12116, Vietnam.\\
$^2$Phenikaa Institute for Advanced Study (PIAS), \\ 
  Phenikaa University, Hanoi 12116, Vietnam. \\
$^3$Institute for Nuclear Science and Technology, VINATOM,\\
 179 Hoang Quoc Viet, Cau Giay, Hanoi 100000, Vietnam.}

\date{Published in Phys. Rev. C {\bf 102}, 045809 (2020)}

\begin{abstract}
The magnetic field of a rotating pulsar might be so strong that the equation of state 
(EOS) of neutron star (NS) matter is significantly affected by the spin polarization 
of baryons. In the present work, the EOS of the spin-polarized nuclear matter is 
investigated in the nonrelativistic Hartree-Fock formalism, using a realistic density-dependent nucleon-nucleon interaction with its spin- and spin-isospin dependence 
accurately adjusted to the Brueckner-Hartree-Fock results for spin-polarized 
nuclear matter. The nuclear symmetry energy and proton fraction are found to increase 
significantly with increasing spin polarization of baryons, leading to a larger 
probability of the direct Urca process in the cooling of magnetar. The EOS of the
$\beta$-stable np$e\mu$ matter obtained at different spin polarization of baryons 
is used as the input for the Tolman-Oppenheimer-Volkoff equations to determine 
the hydrostatic configuration of NS. Based on the GW170817 constraint on the radius 
$R_{1.4}$ of NS with $M\approx 1.4~M_\odot$, our mean-field results show that up 
to 60$\%$ of baryons in the NS merger might be spin-polarized. This result supports 
the magnetar origin of the ``blue" kilonova ejecta of GW170817 suggested by 
Metzger {\it et al.}, and the spin polarization of baryons needs, therefore, 
to be properly treated in the many-body calculation of the EOS of NS matter 
before comparing the calculated NS mass and radius with those constrained 
by the multimessenger GW170817 observation.    
\end{abstract}
\pacs{}
\maketitle

\section{Introduction}\label{sec1} 
The rotating pulsars are usually associated with strong magnetic field ($B$ on 
the order of 10$^{14}$ to 10$^{19}$ G) \cite{Latt07,Bro00,Dex17}, and effects 
of the magnetic field on the equation of state (EOS) of neutron star (NS) matter 
might not be negligible. This important issue has been investigated by many authors 
(see Sec.~9 of Ref. \cite{Latt07} and references therein), and the impact on 
the EOS by the magnetic field of a hydrodynamically stable NS was shown to be 
essential only if the field intensity $B\gtrsim 10^{18}$ G. In particular, 
the complete spin polarization of neutrons likely occurs at $B\gtrsim 4.41\times 
10^{18}$ G \cite{Bro00}. Given a common belief that the magnetic field of NS is 
usually much weaker than the upper limit of $B\approx 10^{19}$ G, it is often 
neglected in numerous mean-field studies of the EOS of NS matter. With the first 
direct observation of the binary NS merger GW170817 by the LIGO and 
Virgo Scientific Collaborations \cite{Abbott17}, a constraint on the tidal 
deformability of NS has been deduced and translated into a constraint on
the radius and mass of NS \cite{Abbott18}. This GW170817 constraint is now 
widely used to validate the mass and radius of NS predicted by different models,
which usually neglect the spin polarization of baryons. 

Recently, the ``blue” kilonova ejecta observed in the aftermath of the NS merger
GW170817 \cite{Abbott17c,Evan17} have been suggested by Metzger {\it et al.} 
\cite{Metzger18} to be caused by both the $\gamma$ decay of the \textit{r}-process nuclei 
and magnetically accelerated wind from the strongly magnetized hypermassive NS 
remnant. A rapidly rotating hypermassive NS remnant having the magnetic field of 
$B\approx (1–3)\times 10^{14}$ G at the surface was found necessary to explain 
the velocity, total mass, and enhanced electron fraction of the blue kilonova ejecta 
\cite{Metzger18}. Because the strength of magnetic field remains quite strong
in the outer core of magnetar \cite{Fuji14}, partial or full spin polarization 
of baryons might well occur during the GW170817 merger.  

In general, the spin polarization of baryons can be explicitly taken into account 
in a microscopic model of nuclear matter (NM) with proper treatment of the 
spin- and spin-isospin dependencies of the in-medium interaction between baryons. 
For example, Vida\~na {\it et al.} \cite{Vidana02,Vidana16} have studied the 
spin-polarized neutron matter within the Brueckner-Hartree-Fock (BHF) formalism 
starting from a free nucleon-nucleon (NN) interaction, to explore the magnetic 
susceptibility of high-density neutron matter and possible phase transition 
to the ferromagnetic state as origin of the NS magnetic field. Aguirre {\it et al.} 
\cite{Agui14} have considered explicitly symmetric NM and neutron matter at finite
temperature embedded in the external magnetic field with $B\simeq 10^{14}$$-$$10^{18}$ G, 
and the spin polarization of baryons $\Delta$ was found strongest at low matter 
densities and becomes weaker with the increasing baryon density $n_{\rm b}$. A 
similar study by Isayev and Yang \cite{Isayev12} shows that high-density neutron 
matter embedded in the strong magnetic field might be partially spin-polarized when 
$B\gtrsim 10^{18}$ G. Tews and Schwenk have recently considered the EOS of
fully spin-polarized NS matter \cite{Tews20}, and concluded that it is ruled out 
by the GW170817 constraint. The complex magnetic-field configuration of the magnetar 
has been investigated by Fujisawa and Kisaka \cite{Fuji14}, and the field intensity 
was shown to diminish gradually to $B\simeq 0$ from the surface to the center of NS, so 
that the baryon matter in the center of NS would not be spin-polarized even for 
magnetar. A partial spin polarization of NS matter is, however, not excluded
by these studies, and it is of interest to investigate its impact on the EOS 
of NS matter. 

Motivated by the magnetar scenario by Metzger {\it et al.} \cite{Metzger18} 
for the ``blue” kilonova ejecta of GW170817, we explore in the present work 
the EOS of the $\beta$-stable np$e\mu$ matter with different partial
spin polarizations of baryons ($0 \lesssim\Delta\lesssim 1$). Like the isospin asymmetry, 
the spin asymmetry of baryons is shown to have a strong impact on the total energy 
and pressure of NM. In particular, the total nuclear symmetry energy has a strong 
repulsive contribution from the spin-symmetry energy, which in turn can affect 
significantly the radius and mass of a magnetar. 

\section{Hartree-Fock approach to the spin-polarized nuclear matter} 
\label{sec2}
The nonrelativistic Hartree-Fock (HF) method \cite{Loan11} is used in the present
work to study the spin-polarized NM at zero temperature, which is characterized 
by the neutron and proton number densities, $n_{\rm n}$ and $n_{\rm p}$, or 
equivalently by the total baryon number density $n_{\rm b}=n_{\rm n}+n_{\rm p}$ 
and neutron-proton asymmetry $\delta=(n_{\rm n}-n_{\rm p})/n_{\rm b}$. The spin 
polarization of baryons is treated explicitly for neutrons and protons by using the 
densities with baryon spin aligned up or down along the magnetic-field axis 
$\Delta_{\rm n,p}=(n_{\uparrow {\rm n,p}}-n_{\downarrow {\rm n,p}})/n_{\rm n,p}$. 
The total HF energy density of NM is obtained as
\begin{equation}
\mathcal{E}=\mathcal{E}_{\rm kin}+{\frac{1}{ 2}}\sum_{k \sigma \tau}
\sum_{k'\sigma '\tau '} [\langle{\bm k}\sigma \tau, {\bm k}' \sigma' \tau'
 |v_{\rm D}|{\bm k}\sigma\tau, {\bm k}' \sigma' \tau' \rangle
+ \langle{\bm k}\sigma \tau, {\bm k}'\sigma' \tau' |v_{\rm EX}|
{\bm k}'\sigma \tau, {\bm k}\sigma' \tau' \rangle], \label{eq1} 
\end{equation}
where $|{\bm k}\sigma \tau\rangle$ are plane waves, and $v_{\rm D}$ and $v_{\rm EX}$ are
the direct and exchange terms of the effective (in-medium) NN interaction. 

We have considered for the present study the density dependent CDM3Y$n$ interaction 
that was successfully used in the HF studies of NM \cite{Loan11,Tan16} and the 
folding model studies of nucleus-nucleus scattering \cite{Kho97,Kho00}. In fact, the 
CDM3Y$n$ interaction is the orginal $G$-matrix based M3Y interaction \cite{Anan83} 
supplemented by the realistic density dependences $F_{\rm st}(n_{\rm b})$ of the 
spin- and spin-isospin-dependent terms of the M3Y interaction  
\begin{align}
& v_{\rm D(EX)}(n_{\rm b}, r)=  F_{00}(n_{\rm b})v_{00}^{\rm D(EX)}(r) + 
 F_{10}(n_{\rm b}) v_{\rm 10}^{\rm D(EX)}(r)({\bm\sigma}\cdot {\bm\sigma}') \nonumber \\
& +F_{01}(n_{\rm b}) v_{01}^{\rm D(EX)}(r)({\bm\tau}\cdot {\bm\tau}')+F_{11}(n_{\rm b}) 
v_{11}^{\rm D(EX)}(r)({\bm\sigma}\cdot {\bm\sigma}') ({\bm\tau}\cdot {\bm\tau}').
\label{eq2}
\end{align}
The radial parts of the direct and exchange terms of the interaction (\ref{eq2}) 
are determined from the spin singlet and triplet components of the M3Y interaction 
\cite{Anan83}, in terms of three Yukawa functions \cite{Kho96} (see Table~\ref{t1}), 
$v^{\rm D(EX)}_{\rm st}(r)=\displaystyle\sum_{\nu=1}^3 Y^{\rm D(EX)}_{\rm st}(\nu)
\exp(-R_\nu r)/(R_\nu r)$. 
\begin{table} [bht]
\caption{Yukawa strengths of the $G$-matrix based M3Y interaction \cite{Anan83}.}
 \vspace{0cm} \label{t1}
\addtolength{\tabcolsep}{3pt}
\begin{tabular}{cccccccccc}\hline\hline 
$\nu$ & $R_{\nu}$ & $Y^{\rm D}_{\rm 00}(\nu)$ & $Y^{\rm D}_{\rm 10}(\nu)$ & 
$Y^{\rm D}_{\rm 01}(\nu)$ & $Y^{\rm D}_{\rm 11}(\nu)$ & $Y^{\rm EX}_{\rm 00}(\nu)$ & 
$Y^{\rm EX}_{\rm 10}(\nu)$ & $Y^{\rm EX}_{\rm 01}(\nu)$ & $Y^{\rm EX}_{\rm 11}(\nu)$ \\
& (fm$^{-1}$) & (MeV) & (MeV) & (MeV) & (MeV) & (MeV) & (MeV) & (MeV) & (MeV)  \\ \hline
1 & 4.0 & 11061.625 & 938.875 & 313.625 & -969.125& -1524.25 & -3492.75 & -4118.0 & -2210.0 \\
2 & 2.5 & -2537.5   & -36.0   & 223.5 & 450.0 & -518.75 & 795.25 & 1054.75 & 568.75  \\
3 & 0.7072 & 0.0    & 0.0     & 0.0 &   3.4877 & -7.8474 & 2.6157 & 2.6157 & -0.8719 \\ 
\hline\hline
\end{tabular} 
\end{table}

Then the total-energy density (\ref{eq1}) can be obtained as
\begin{equation}
\mathcal{E}=\frac{3}{10}\sum_{\sigma\tau}\frac{\hbar^2 k^2_{F\sigma\tau}}
 {m_\tau}n_{\sigma \tau}+F_{00}(n_{\rm b})\mathcal{E}_{00} 
+ F_{10}(n_{\rm b})\mathcal{E}_{10} + F_{01}(n_{\rm b})\mathcal{E}_{01}
+ F_{11}(n_{\rm b})\mathcal{E}_{11} \label{eq3} 
\end{equation} 
where $\sigma=\uparrow,\downarrow$ and $\tau={\rm n},{\rm p}$. The potential-energy 
density of NM is determined by using
\begin{align}
& \mathcal{E}_{00}=\frac{1}{2}\left[n_{\rm b}^2 J^D_{00}+ 
 \int A^2_{00} v^{\rm EX}_{00}(r) d^3r\right], \nonumber\\
& \mathcal{E}_{10}=\frac{1}{2}\left[n_{\rm b}^2 J^D_{10}
 \left(\Delta_{\rm n}\frac{1+\delta}{2}+\Delta_{\rm p}\frac{1-\delta}{2}\right)^2 
 + \int A^2_{10} v^{\rm EX}_{10}(r) d^3r\right], \nonumber \\
& \mathcal{E}_{01}=\frac{1}{2}\left[n_{\rm b}^2 J^D_{01}\delta^2 + 
 \int A^2_{01} v^{\rm EX}_{01}(r) d^3r\right], \nonumber \\
& \mathcal{E}_{11}=\frac{1}{2}\left[n_{\rm b}^2 J^D_{11}
 \left(\Delta_{\rm n}\frac{1+\delta}{2}-\Delta_{\rm p}\frac{1-\delta}{2}\right)^2
 + \int A^2_{\rm 11} v^{\rm EX}_{11}(r) d^3r\right]. \label{eq4} 
\end{align}
$J^D_{\rm st}=\displaystyle\int v_{\rm st}(r) d^3r$ is the volume integral
of the direct interaction, and the exchange integrals in Eq. (\ref{eq4}) are 
evaluated with
\begin{align}
& A_{00}=n_{\uparrow {\rm n}} \widehat{j}_1(k_{F_{\uparrow {\rm n}}}r)+
 n_{\downarrow {\rm n}} \widehat{j}_1(k_{F_{\downarrow {\rm n}}}r)+
 n_{\uparrow {\rm p}} \widehat{j}_1(k_{F_{\uparrow {\rm p}}}r)+
 n_{\downarrow {\rm p}} \widehat{j}_1(k_{F_{\downarrow {\rm p}}}r) \nonumber\\
& A_{10}=n_{\uparrow {\rm n}} \widehat{j}_1(k_{F_{\uparrow {\rm n}}}r)-
 n_{\downarrow {\rm n}} \widehat{j}_1(k_{F_{\downarrow {\rm n}}}r)+
 n_{\uparrow {\rm p}} \widehat{j}_1(k_{F_{\uparrow {\rm p}}}r)-
 n_{\downarrow {\rm p}} \widehat{j}_1(k_{F_{\downarrow {\rm p}}}r) \nonumber\\
& A_{01}=n_{\uparrow {\rm n}} \widehat{j}_1(k_{F_{\uparrow {\rm n}}}r)+
 n_{\downarrow {\rm n}} \widehat{j}_1(k_{F_{\downarrow {\rm n}}}r)-
 n_{\uparrow {\rm p}} \widehat{j}_1(k_{F_{\uparrow {\rm p}}}r)-
 n_{\downarrow {\rm p}} \widehat{j}_1(k_{F_{\downarrow {\rm p}}}r) \nonumber\\
& A_{11}=n_{\uparrow {\rm n}} \widehat{j}_1(k_{F_{\uparrow {\rm n}}}r)-
 n_{\downarrow {\rm n}} \widehat{j}_1(k_{F_{\downarrow {\rm n}}}r)-
 n_{\uparrow {\rm p}} \widehat{j}_1(k_{F_{\uparrow {\rm p}}}r)+
 n_{\downarrow {\rm p}} \widehat{j}_1(k_{F_{\downarrow {\rm p}}}r), \label{eq5}
\end{align}
where $\hat{j}_1(x)=3j_1(x)/x$, and $j_1(x)$ is the first-order spherical Bessel 
function. The Fermi momentum of the spin-polarized baryon is determined as 
$k_{F\sigma\tau}=(6\pi^2n_{\sigma\tau})^{1/3}$. 
   
\begin{figure}[bht]\vspace{-1cm}
\includegraphics[angle=0,width=1.1\textwidth]{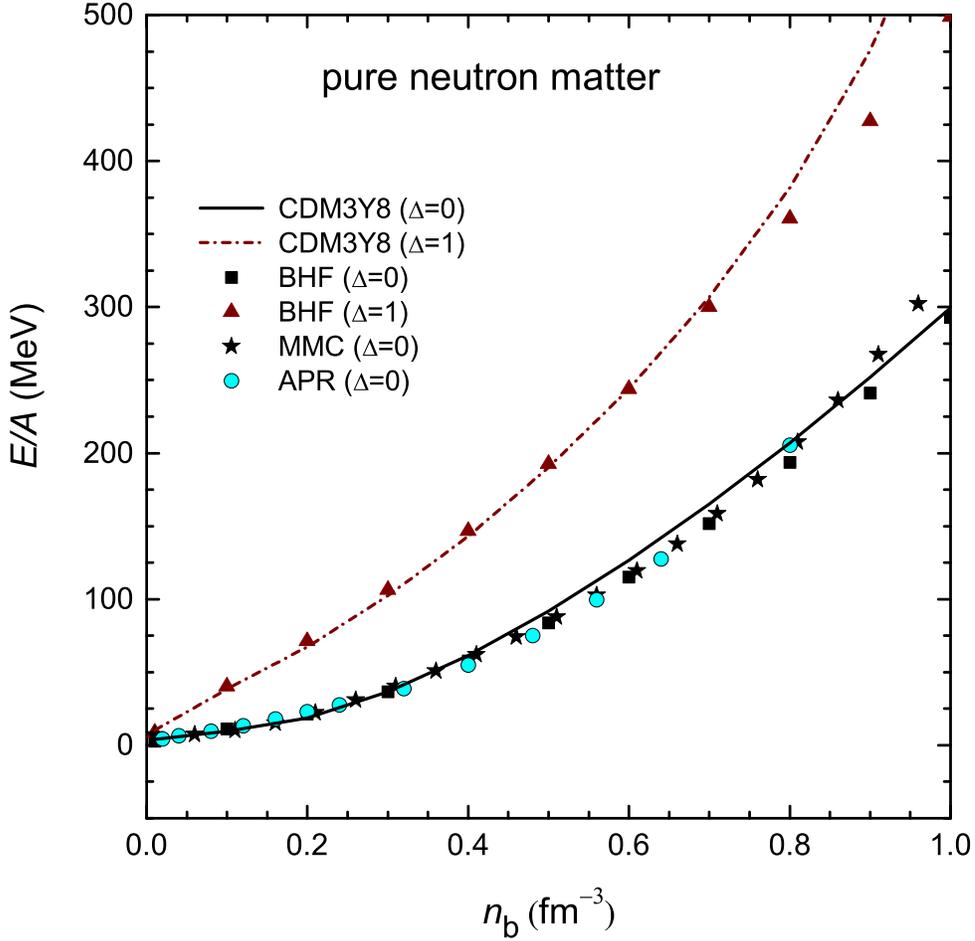}\vspace{-4cm}
 \caption{Energy per baryon of pure neutron matter with the neutron spin polarization 
$\Delta=0$ and 1 given by the HF calculation (\ref{eq5}) using the newly parametrized 
CDM3Y8 interaction, in comparison with results of the BHF calculation (squares and
triangles) \cite{Vidana16}. The circles and stars are results of the \textit{ab-initio} calculations 
by Akmal, Pandharipande, and Ravenhall (APR) \cite{APR} and microscopic Monte Carlo (MMC) 
calculation by Gandolfi {\it et al.} \cite{MMC}, respectively.} 
\label{f1}
\end{figure}  
One can see that the spin polarization of baryons gives rise to the nonzero
contribution from both the $\mathcal{E}_{10}$ and $\mathcal{E}_{11}$ terms to the total 
NM energy density (\ref{eq3}). Therefore, the density dependencies $F_{10}(n_{\rm b})$ 
and $F_{11}(n_{\rm b})$ of the CDM3Y$n$ interaction (\ref{eq2}) need to be properly 
determined for the present HF study. The spin-isospin independent (isoscalar) density 
dependence $F_{00}(n_{\rm b})$ was parametrized \cite{Kho97} to correctly reproduce 
the saturation properties of symmetric NM at the baryon density $n_0\approx 0.16$ fm$^{-3}$, 
and the isospin dependent (isovector) density dependence $F_{01}(n_{\rm b})$ was adjusted 
to the BHF results of NM and fine tuned in the coupled-channel study of the charge 
exchange $(p,n)$ reaction to isobar analog states in finite nuclei \cite{Kho07,Kho14}. 
In the present work we have parametrized the density dependencies $F_{10}(n_{\rm b})$ and 
$F_{11}(n_{\rm b})$ of the spin- and spin-isospin-dependent parts of the CDM3Y$n$ 
interaction in the same functional form as that used earlier for $F_{00}(n_{\rm b})$ 
and $F_{01}(n_{\rm b})$, and the parameters were adjusted to obtain the HF results 
for the spin-polarized neutron matter close to those of the BHF calculation by 
Vida\~na {\it et al.} \cite{Vidana16} using the Argonne V18 free NN potential added 
by the Urbana IX three-body force. The parameters of $F_{00}(n_{\rm b})$ and 
$F_{01}(n_{\rm b})$ were also slightly readjusted for a better agreement of the HF 
results with those of the \textit{ab-initio} calculations \cite{APR,MMC} at high baryon densities 
(see Fig.~\ref{f1}). This new version of the CDM3Y$n$ interaction is referred to 
hereinafter as the CDM3Y8 interaction, with all parameters of the density dependence 
given explicitly in Table~\ref{t2}. 
\begin{table} [bht]
\caption{Parameters of the density dependence of the CDM3Y8  interaction (\ref{eq2}),\\
 \centerline{$F_{\rm st}(n_{\rm b})=C_{\rm st}[1+\alpha_{\rm st}\exp(-\beta_{\rm st}n_{\rm b})
 +\gamma_{\rm st}n_{\rm b}]$.}}
\vspace{0cm} \label{t2}
\addtolength{\tabcolsep}{3pt}
\begin{tabular}{ccccccc}\\ \hline\hline
$\rm st$ & $C_{\rm st}$& $\alpha_{\rm st}$& $\beta_{\rm st}$ & $\gamma_{\rm st}$ \\
 &  &  & (fm$^3$) & (fm$^3$)  \\ \hline
00 & 0.2658 & 3.8033 & 1.4099  &-4.300  \\ 
01 & 0.2643 & 6.3836 & 10.2566 & 6.3549 \\    
10 & 0.2161 & 3.7510 & -3.3396 & 9.9329 \\    
11 & 0.7572 & 1.9967 & 33.2012 & 0.2989 \\ \hline\hline    
\end{tabular}
\end{table}

\section{Nuclear symmetry energy}
 \label{sec3}
\begin{figure}[bht!]\vspace{-0.5cm}\hspace*{-1cm}
\includegraphics[angle=0,width=1.2\textwidth]{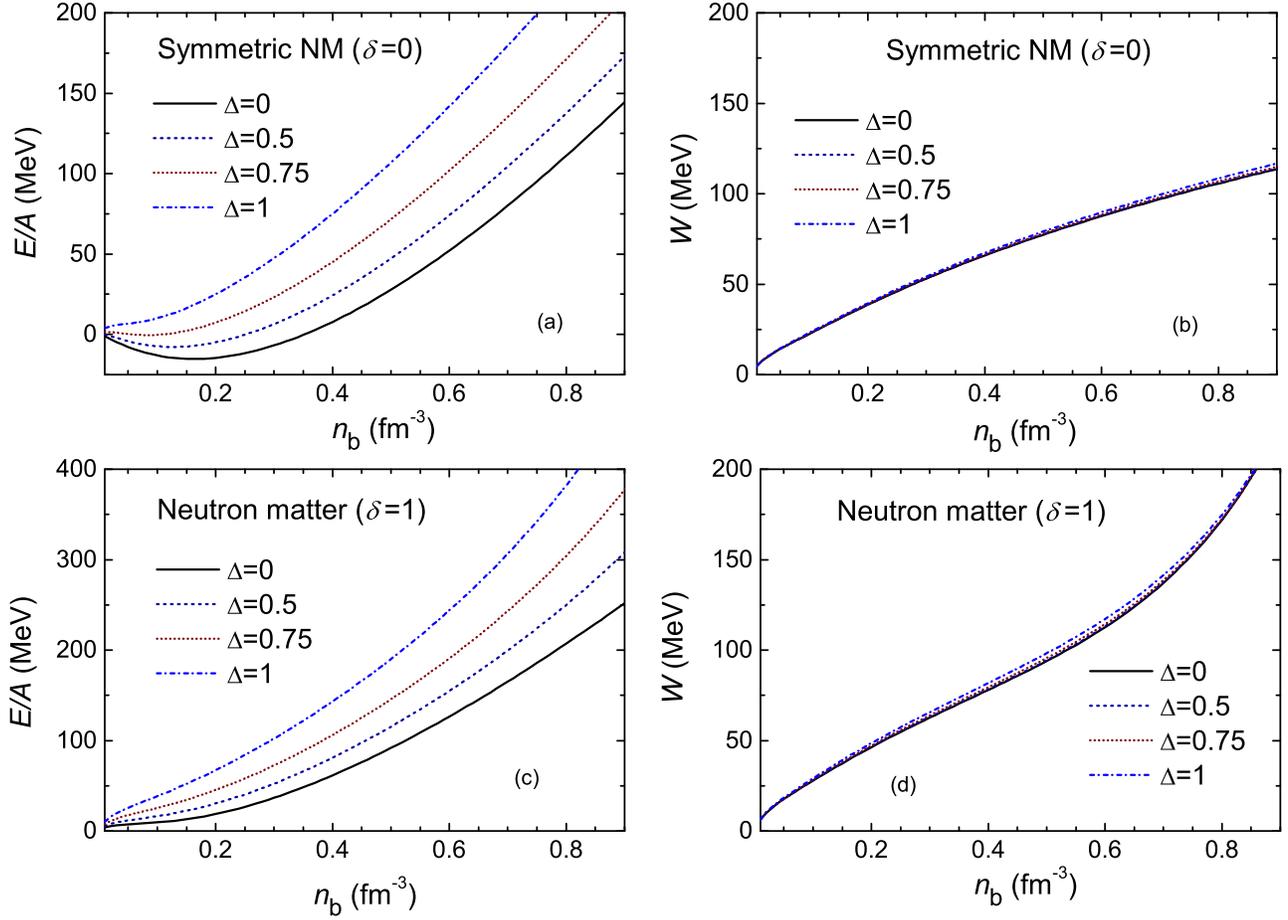}\vspace{0cm}
 \caption{The present HF results obtained at different spin polarizations 
 $\Delta$ of baryons for the energy per baryon $E/A$  and spin-symmetry energy $W$ 
 of symmetric NM, panels (a) and (b), respectively,  and those of neutron matter, 
 panels (c) and (d), respectively.} \label{f2}
\end{figure}

Although the neutron and proton magnetic moments are of different strengths and 
of opposite signs, in the presence of strong magnetic field $|\Delta_n|$ and 
$|\Delta_p|$ should be of the same order. We have assumed, for simplicity, 
the baryon spin polarization $\Delta=\Delta_n\approx -\Delta_p$ in the present 
HF study. The total NM energy per baryon $E/A$ is then obtained in the 
{\it isospin} symmetry as
\begin{equation}
\frac{\mathcal{E}}{n_{\rm b}}\equiv\frac{E}{A}(n_{\rm b},\Delta,\delta)=
\frac{E}{A}(n_{\rm b},\Delta,\delta=0)+ S(n_{\rm b},\Delta)\delta^2+O(\delta^4)+... 
\label{eq6}
\end{equation}
The contribution from $O(\delta^4)$ and higher-order terms in Eq.~(\ref{eq6}) is 
small and neglected in the \emph{parabolic} approximation \cite{Kho96}, where the 
isospin-symmetry energy $S(n_{\rm b},\Delta)$ equals the energy required per baryon 
to change symmetric NM into the pure neutron matter. Governed by the same SU(2) 
symmetry, such a parabolic approximation is also valid for the {\it spin} symmetry, 
and the total NM energy per baryon can be alternatively obtained as 
\begin{equation}
\frac{\mathcal{E}}{n_{\rm b}}\equiv\frac{E}{A}(n_{\rm b},\Delta,\delta)=
\frac{E}{A}(n_{\rm b},\Delta=0,\delta)+ W(n_{\rm b},\delta)\Delta^2+O(\Delta^4)+... 
\label{eq7}
\end{equation}
The exact spin-symmetry energy $W$ given by the HF calculation (\ref{eq7}) of 
symmetric NM and neutron matter are shown on the right panel of Fig.~\ref{f2}, 
and one can see that $W$ is approximately $\Delta$-independent, and the  
contribution from $O(\Delta^4)$ and higher-order terms to the NM energy 
(\ref{eq7}) is indeed negligible.   

Given the quadratic dependence of the NM energy on the spin polarization of baryons
and positive strength of the spin-symmetry energy $W$ over the whole range of densities,
it is sufficient to consider only $\Delta\geqslant 0$ in the present HF study. 
One can see in the left panel of Fig.~\ref{f2} that the nonzero spin polarization 
significantly stiffens the EOS of NM. In particular, the symmetric NM becomes unbound 
by the strong interaction at $\Delta\gtrsim 0.75$, in about the same way as the 
asymmetric NM becomes unbound by the in-medium NN interaction with the increasing 
neutron-proton asymmetry $\delta\gtrsim 0.75$ \cite{Kho96}.  
\begin{figure}[bht]\vspace{-1.5cm}
\includegraphics[angle=0,width=1.1\textwidth]{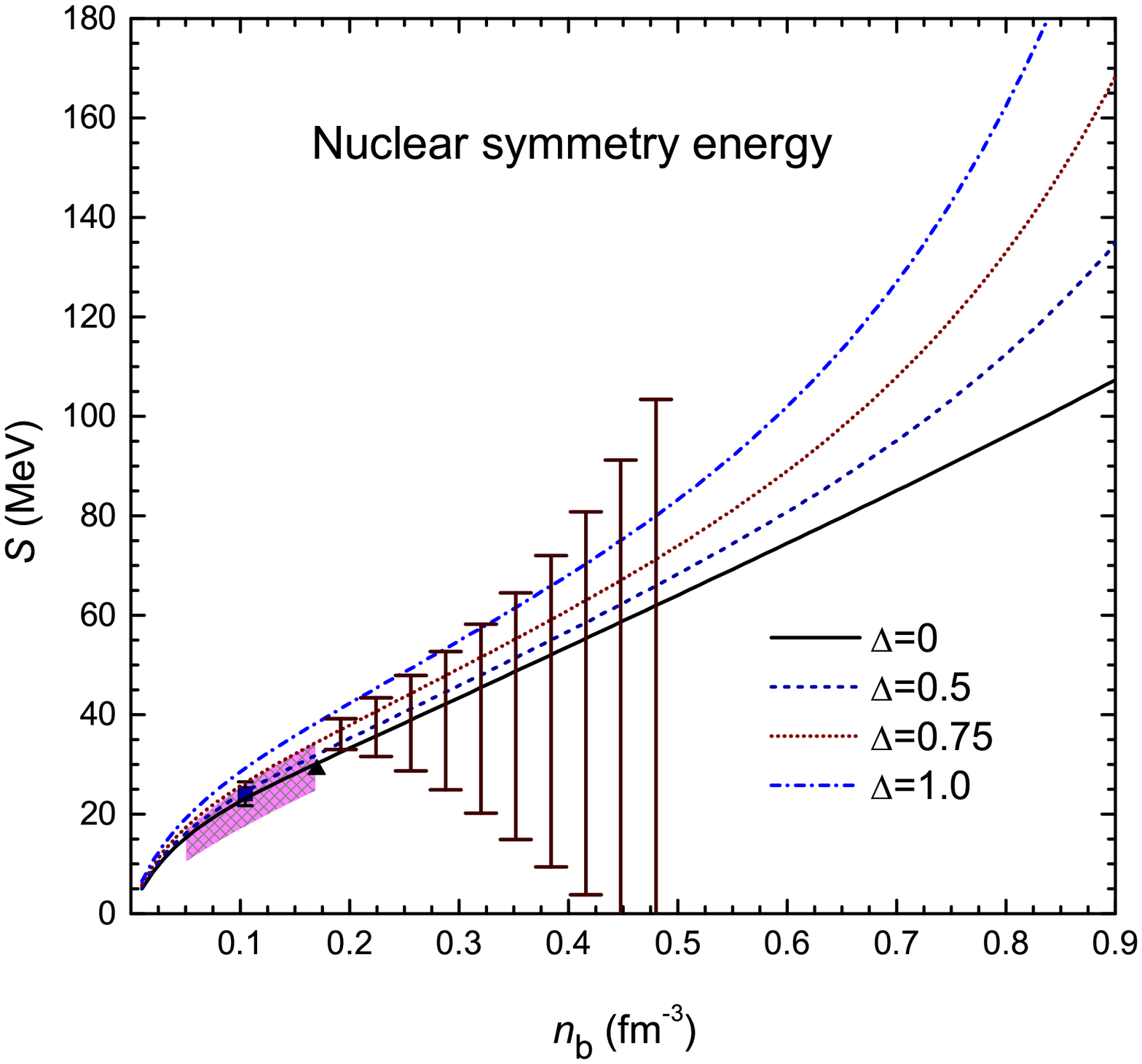}\vspace{-2.5cm}
 \caption{The nuclear symmetry energy $S(n_{\rm b},\Delta)$ given by the HF calculation 
(\ref{eq6}) assuming different spin polarizations of baryons $\Delta$. The shaded 
region is the range constrained by the data of HI collisions \cite{Tsang11,Ono03}. 
The square and triangle are values suggested by the nuclear structure studies 
\cite{Tri08,Fur02}. The vertical bars are the empirical range obtained at the 90\% 
confidence level in a statistical Bayesian analysis \cite{Xie19} of the NS radius 
$R_{1.4}$ versus the GW170817 constraint \cite{Abbott18}.} \label{f3} 
\end{figure} 
 
The isospin-symmetry energy $S(n_{\rm b},\Delta)$, widely discussed in the literature 
as the {\bf nuclear symmetry energy}, is a key characteristics of the EOS of neutron-rich 
NM. In particular, the knowledge about the density dependence of $S(n_{\rm b})$ is 
extremely important for the determination of the nuclear EOS and it has been, 
therefore, a longstanding goal of numerous nuclear physics and nuclear astrophysics 
studies (see, e.g., Refs.~\cite{BaLi08,Hor14,Latt14}). However, the results of these 
studies were mainly obtained for the spin-saturated NM, and describe, therefore, 
the behavior of $S(n_{\rm b},\Delta=0)$.  

The nuclear symmetry energy is rather well constrained at low baryon densities by 
the analyses of the (isospin dependent) data of heavy-ion (HI) collisions 
\cite{Tsang11,Ono03} as well as the structure studies of the giant dipole resonance 
\cite{Tri08} or neutron skin \cite{Fur02}. Our HF results for $S(n_{\rm b},\Delta)$ 
are compared with the empirical data in Fig.~\ref{f3}, and a significant increase 
of the nuclear symmetry energy is found with the increasing spin polarization 
of baryons $\Delta$. At low densities, the calculated $S(n_{\rm b},\Delta)$ values 
fall within the empirical range when the baryon spin polarization 
$\Delta\lesssim 0.75$. The behavior of the nuclear symmetry energy at high baryon 
densities ($n_{\rm b}>n_0$) remains not well determined. However, the mass and radius 
of NS (given, e.g., by the Tolman-Oppenheimer-Volkoff equations using different 
EOS's of NS matter) are proven to be strongly sensitive to the strength and slope 
of $S(n_{\rm b})$ at high densities \cite{Loan11,Xie19}. Recently, Xie and Li 
\cite{Xie19} have inferred the symmetry energy at baryon densities up to $3n_0$ 
from a statistical Bayesian analysis of the correlation of different EOS's of the 
np$e\mu$ matter and associated radius $R_{1.4}$ of NS with mass $M\approx 1.4~M_{\odot}$ 
versus the GW170817 constraint on $R_{1.4}$ imposed by the NS tidal deformability. 
In particular, the EOS with the symmetry energy at twice the saturation density 
$S(2n_0)\approx 40-60$ MeV give $R_{1.4}$ radii within the range constrained 
by the tidal deformability. 
The empirical $S(n_{\rm b})$ values suggested for baryon densities up to $3 n_0$ 
at 90\% confidence level \cite{Xie19} are shown in Fig.~\ref{f3}, and they cover 
the symmetry energy predicted by the HF calculation of neutron-rich NM over the whole 
range of the spin polarization of baryons $0\lesssim\Delta\lesssim 1$. One can 
trace in Fig.~\ref{f3} that the symmetry energy obtained with a narrower uncertainty 
(at 68 \% confidence level) $S(2n_0)\approx 39.2^{+12.1}_{-8.2}$ MeV \cite{Xie19}
also covers all possible spin polarizations. 

To explore the density dependence of the nuclear symmetry energy, $S$ is often 
expanded around the saturation density $n_0$ \cite{BaLi08,Hor14,Latt14} in terms 
of the symmetry coefficient $J$, slope $L$ and curvature $K_{\rm sym}$. With the spin 
polarization of baryons treated explicitly in the present HF study, these quantities
are now dependent on the spin polarization $\Delta$, and we obtain 
\begin{equation}
 S(n_{\rm b},\Delta)=J(\Delta)+\frac{L(\Delta)}{3}\left(\frac{n_{\rm b}-n_0}{n_0}\right) +
 \frac{K_{\rm sym}(\Delta)}{18}\left(\frac{n_{\rm b}-n_0}{n_0}\right)^2+ ... \label{eq8}
\end{equation}
$J(\Delta)$, $L(\Delta)$, $K_{\rm sym}(\Delta)$, and the incompressibility $K_0(\Delta)$ 
of symmetric NM at the saturation density (which also depends on the spin polarization 
of baryons) are the most important characteristics of the EOS of the spin-polarized 
NM. The $J,L,K_{\rm sym},K_0$ values given by the present HF calculation using the 
CDM3Y8 interaction are given in Table~\ref{t3}. 
\begin{table*} [bht]
\caption{The symmetry coefficient $J$, slope $L$, and curvature $K_{\rm sym}$ 
of the symmetry energy (\ref{eq8}), and incompressibility $K_0$ of symmetric NM 
at the saturation density $n_0$ given by the HF calculation of the spin-polarized 
NM using the CDM3Y8 interaction.} \vspace{0.5cm} \label{t3}
\addtolength{\tabcolsep}{3pt}
\begin{tabular}{lcccc}\hline \hline
$\Delta$ & $J$ & $L$ & $K_{\rm sym}$ & $K_0$  \\
  &  (MeV)  & (MeV) & (MeV) & (MeV) \\ \hline
 0.0 &  29.5 & 50.6 & -254 & 244  \\
 0.1 &  29.6 & 50.7 & -256 & 243 \\
 0.2 &  29.8 & 51.0 & -264 & 242 \\
 0.3 &  30.1 & 51.6 & -275 & 240 \\
 0.4 &  30.6 & 52.5 & -291 & 238\\
 0.5 &  31.3 & 53.5 & -314 & 234 \\
 0.6 &  32.1 & 54.9 & -340 & 230 \\ 
 0.7 &  33.1 & 56.6 & -369 & 225 \\
 0.8 &  34.3 & 58.7 & -402 & 219 \\
 0.9 &  35.7 & 61.3 & -450 & 213 \\ 
 1.0 &  37.5 & 64.7 & -505 & 205 \\ \hline  \hline
\end{tabular} \vspace{0.cm}        
\end{table*}
Among these quantities, the incompressibility $K_0$ of symmetric NM has been the 
key research topic of numerous structure studies of nuclear monopole excitations 
(see, e.g., review \cite{Garg18} and references therein) as well as the studies 
of HI collisions and refractive nucleus-nucleus scattering \cite{Kho07r}. These
researches have pinned down this quantity to $K_0\approx 240\pm 20$ MeV. 
The symmetry coefficient and slope of the nuclear symmetry energy (\ref{eq8}) 
were also extensively investigated and inferred independently from different 
analyses of terrestrial nuclear physics experiments and astrophysical observations, 
and they are now constrained to $J\approx 31.7 \pm 3.2$ MeV and 
$L\approx 58.7\pm 28.1$ MeV  \cite{Zhang18}. The $K_{\rm sym}$ value is still not 
well determined, and remains within a wide range 
$-400$ MeV $\lesssim K_{\rm sym}\lesssim 100$ MeV \cite{Zhang18}. The HF
results for $J, L, K_{\rm sym}$, and $K_0$ of the spin-saturated NM with 
$\Delta=0$ agree well with the empirical values, and remain within the empirical 
boundaries with the spin polarization of baryons $0\lesssim\Delta\lesssim 0.8$.    

\section{Beta-stable neutron star matter}
\label{sec4}
For the EOS of inhomogeneous NS crust, we have adopted that given by the nuclear 
energy-density functional (EDF) calculation \cite{Chamel18,Chamel19} using the 
BSk24 Skyrme functional, with atoms being fully ionized and electrons forming 
a degenerate Fermi gas. At the edge density $n_{\rm edge}\approx 0.076$ fm$^{-3}$, 
a weak first-order phase transition takes place between the NS crust and uniform 
core of NS. At baryon densities $n_{\rm b}\gtrsim n_{\rm edge}$ the NS core is 
described as a homogeneous matter of neutrons, protons, electrons, and negative 
muons ($\mu^-$ appear at $n_{\rm b}$ above the muon threshold density 
$\mu_e>m_\mu c^2\approx 105.6$ MeV). 

\subsection*{Density-independent spin polarization of the np$e\mu$ matter} 
To explore the impact of the spin polarization of baryons on the EOS of the 
$\beta$-stable np$e\mu$ matter of NS, we have assumed for simplicity that the spin 
polarization of baryons $\Delta$ is density independent, and varied $\Delta$ 
within the range $(0\to 1)$ at each considered density, as done above for the 
spin-polarized NM. Then, the total-energy density $\mathcal{E}$ of the np$e\mu$ matter 
(including the rest energy) is determined as
\begin{equation}
\mathcal{E}(n_{\rm n},n_{\rm p},n_e,n_\mu,\Delta)=
\mathcal{E}_{\rm HF}(n_{\rm n},n_{\rm p},\Delta)+n_{\rm n}m_{\rm n}c^2
+n_{\rm p}m_{\rm p}c^2+\mathcal{E}_e(n_e)+\mathcal{E}_\mu(n_\mu) \label{eq9}, 
\end{equation}
where $\mathcal{E}_{\rm HF}(n_{\rm n},n_{\rm p},\Delta)$ is the HF
energy density (\ref{eq3}) of the spin-polarized NM, $\mathcal{E}_e$ and 
$\mathcal{E}_\mu$ are the energy densities of electrons and muons given 
by the relativistic Fermi gas model \cite{Dou01}. In such a Fermi gas model,
the spin polarization of leptons does not affect the total energy density 
$\mathcal{E}$, and the lepton number densities $n_e$ and $n_\mu$ can be determined 
from the charge neutrality condition ($n_{\rm p}=n_e+n_\mu$) and the $\beta$-equilibrium 
of (neutrino-free) NS matter in the same way as done for the spin unpolarized
NS matter (see Ref.~\cite{Loan11} for more details).

\begin{figure}[bht]\vspace{0cm}
\includegraphics[angle=0,width=0.9\textwidth]{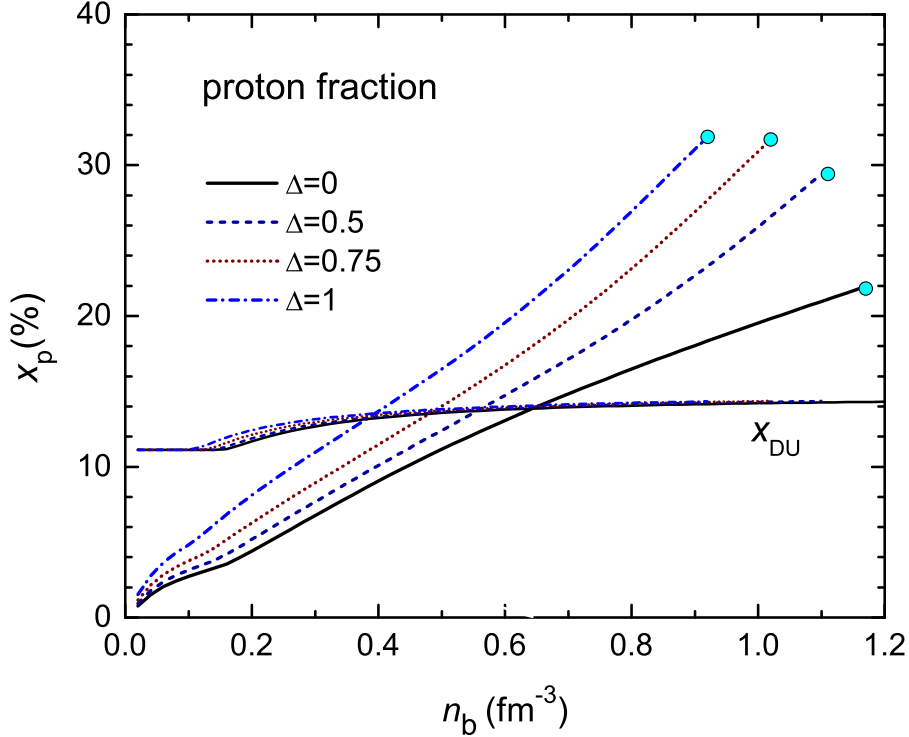}\vspace{-0.5cm}
 \caption{The proton fraction $x_{\rm p}$ determined from the charge neutrality 
of the spin-polarized $\beta$-stable np$e\mu$ matter obtained with the HF energy 
density (\ref{eq3}) using the CDM3Y8 interaction. The circles are $n_{\rm p}$ values 
calculated at the maximum central densities $n_{\rm c}$, and the thin lines are the 
DU thresholds (\ref{eq10}).} \label{f4} 
\end{figure}

The density-dependent proton fraction $x_{\rm p}(n_{\rm b})$ is a key input for 
the determination of the NS cooling rate. In particular, the direct Urca (DU) 
process of NS cooling via neutrino emission is possible only if the proton 
fraction is above the DU threshold $x_{\rm DU}$ \cite{Loan11}: 
\begin{equation}
 x_{\rm DU}(n_{\rm b})=\frac{1}{1+\left[1+r_e^{1/3}(n_{\rm b})\right]^3}, 
 \label {eq10}
\end{equation}
where $r_e(n_{\rm b})=n_e/(n_e+n_\mu)$ is the leptonic electron fraction at the given
baryon number density. At low densities $r_e=1$, and $x_{\rm DU}\approx 11.1\%$,  
which corresponds to the muon-free threshold for the DU process. Because the lepton-baryon 
interaction is neglected in the present study, the $x_{\rm DU}$ value determined from 
the $\beta$-equilibrium condition depends very weakly on the spin-polarization 
of baryons. The proton fraction $x_{\rm p}$ of the spin-polarized $\beta$-stable 
np$e\mu$ matter obtained with the HF energy density (\ref{eq3}) using the CDM3Y8 
interaction is shown in Fig.~\ref{f4}, and one can see that $x_{\rm p}$ increases   
significantly with the increasing spin polarization of baryons, and it exceeds the
DU threshold at densities $n_{\rm b}\gtrsim 2n_0$ if baryons are completely spin 
polarized ($\Delta=1$). From the behavior of $n_{\rm p}$ shown in Fig.~\ref{f4} 
we find that the electron fraction in the $\beta$-stable np$e\mu$ matter
also increases with the increasing $\Delta$ and might reach up to 20\%$-$30\% at 
high densities when $\Delta$ approaches 1. It is remarkable that such a high 
electron fraction was found in the blue kilonova ejecta following the NS 
merger GW170817 \cite{Abbott17c,Evan17}, and suggested by Metzger \textit{et al.} 
\cite{Metzger18} to be of the magnetar origin. 

\begin{table} [bht]
\caption{Configuration of NS given by the EOS of the spin-polarized $\beta$-stable 
np$e\mu$ matter obtained with the CDM3Y8 interaction. $M_{\rm max}$ and $R_{\rm max}$ 
are the maximal gravitational mass and radius; $R_{1.4}$, $n_c$, and $P_c$ are 
the radius of the NS with $M\approx 1.4~M_\odot$, central baryon number density, and 
central total pressure, respectively. $P(2n_0)$ is the total pressure at twice the 
saturation density.} \vspace{0.5cm} \label{t4}
\begin{tabular}{ccccccc}\hline\hline
$\Delta$ & $M_{\rm max}$ & $R_{\rm max}$ &$R_{1.4}$& $n_c$ & $P_c$ & $P(2n_0)$  \\
 &($M_\odot$)&(km) & (km)& (fm$^{-3}$)&($10^{35}$dyn/cm$^2$)& ($10^{34}$dyn/cm$^2$)\\ \hline
0.0 & 1.98 & 10.3 & 12.0 & 1.17 & 9.9 & 3.4    \\
0.1 & 1.99 & 10.3 & 12.0 & 1.16 & 10.0 & 3.5    \\
0.2 & 2.00 & 10.4 & 12.1 & 1.16 & 10.2 & 3.6    \\ 
0.3 & 2.02 & 10.4 & 12.3 & 1.15 & 10.4 & 3.6   \\
0.4 & 2.04 & 10.5 & 12.5 & 1.14 & 10.9 & 3.9    \\
0.5 & 2.06 & 10.6 & 12.8 & 1.11 & 10.3 & 4.1   \\ 
0.6 & 2.08 & 10.8 & 13.1 & 1.08 & 10.0 & 4.4    \\
0.7 & 2.10 & 11.1 & 13.6 & 1.04 &  9.5 & 4.6   \\
0.8 & 2.12 & 11.3 & 14.1 & 1.01 &  9.0 & 5.0  \\
0.9 & 2.14 & 11.7 & 14.8 & 0.96 &  8.1 & 5.4  \\
1.0 & 2.16 & 12.0 & 15.6 & 0.92 &  7.7 &  5.8  \\ \hline \hline
\end{tabular} \vspace{0.cm}        
\end{table}
\begin{figure}[bht]\vspace{-0.5cm}
\includegraphics[angle=0,width=0.8\textwidth]{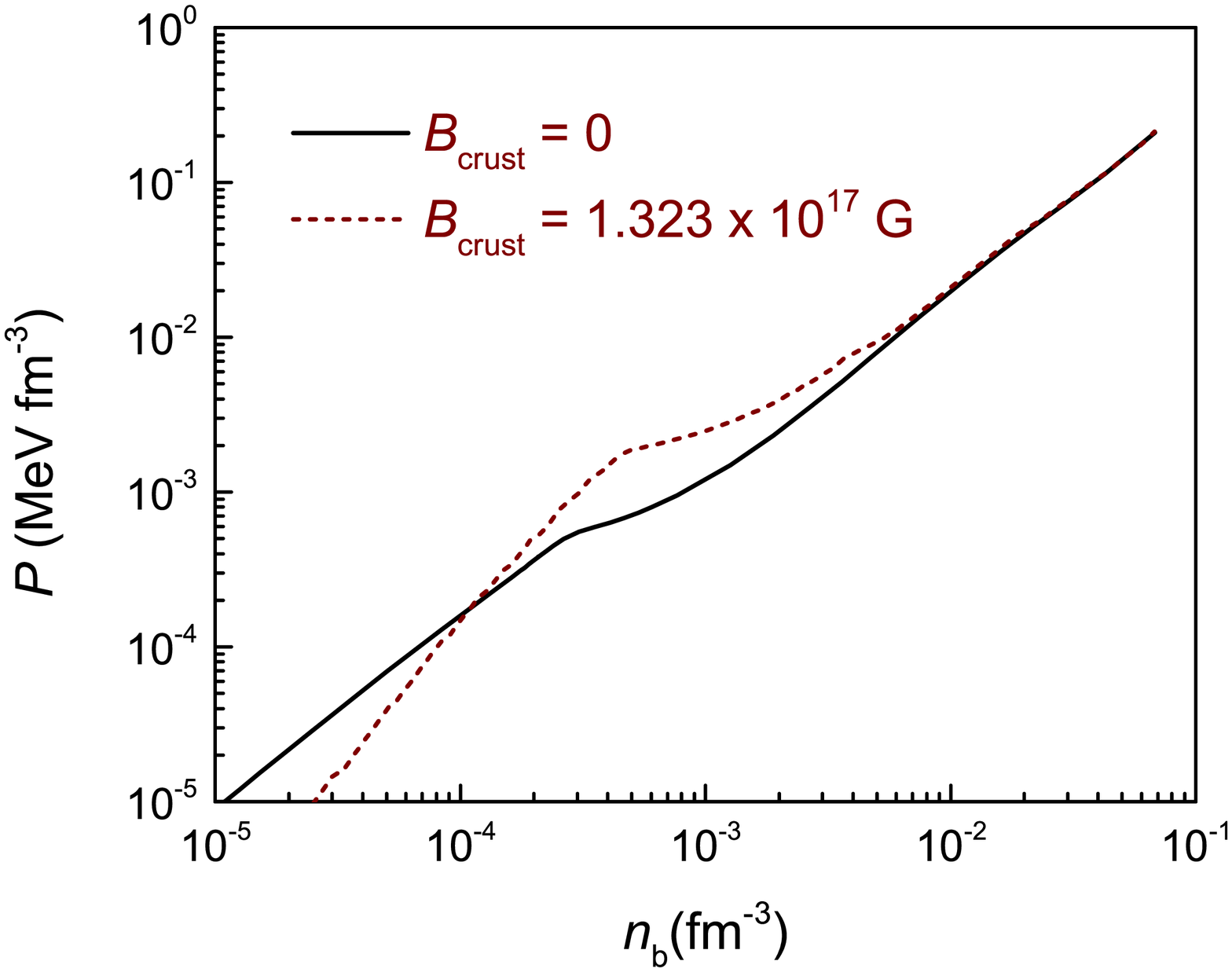}\vspace{-0.5cm}
 \caption{Pressure $P$ in the unmagnetized ($B_{\rm crust}=0$) and magnetized 
($B_{\rm crust}\neq 0$) crust of NS given by the EDF calculation 
\cite{Chamel18,Chamel19} using the BSk24 Skyrme functional.} \label{f5} 
\end{figure} 

The mass density $\rho$ and total pressure $P$ of NS crust given by the EDF 
calculation at baryon densities below the edge density 
$n_{\rm edge}\approx 0.076$ fm$^{-3}$, and those of the uniform and spin-polarized 
np$e\mu$ matter given by the HF calculation at $n_{\rm b}\gtrsim n_{\rm edge}$ 
have been used as inputs for the Tolman-Oppenheimer-Volkoff (TOV) equations 
to determine the hydrostatic configuration of NS (see Table~\ref{t4}). 
For a consistent mean-field study of the spin-polarized np$e\mu$ matter of NS, 
we have used in this work two versions of NS crust: the unmagnetized crust 
($B_{\rm crust}=0$) and crust embedded in the magnetic field of 
$B_{\rm crust}=1.323\times 10^{17}$ G \cite{Chamel19}. The effects of magnetic 
field on the EOS and the composition of NS crust were shown mainly due to the 
Landau quantization of the electron motion, with most protons and neutrons 
remaining ``packed" in nuclei with $Z\approx 40\sim 50$ inside in the Wigner-Seitz cell 
\cite{Chamel19}. The impact of the magnetic field with $B_{\rm crust}\sim 10^{17}$ G 
is strong in the outer crust (see Fig.~\ref{f5}), while the EOS of the inner crust 
remains almost unchanged compared to that of the unmagnetized crust at baryon 
densities $n_{\rm b}\gtrsim 0.01$ fm$^{-3}$.   
As a result, the main properties of NS shown in Table~\ref{t4} are not affected
by the magnetization of the NS crust. We note that the spin polarization of free 
baryons and spin-unsaturated nuclei inside the Wigner-Seitz cell has not 
been taken into account in the EDF approach \cite{Chamel19}, and it might result 
in a stronger impact of the magnetic field on the EOS of the NS crust.  

Because NS matter becomes less compressible (see $K_0$ values in Table~\ref{t3})
when the spin polarization of baryons is nonzero, the central density $n_c$ and pressure 
$P_c$ decrease with increasing $\Delta$. As a result, the NS expands its size and 
the maximal gravitational mass $M_{\rm max}$ and radius $R_{\rm max}$ become larger 
with increasing spin polarization of baryons. The LIGO and Virgo data of the NS
merger GW170817 were analyzed by Abbott {\it et al.} \cite{Abbott18} to put constraints 
on the tidal deformability of two merging neutron stars, which were then translated into 
constraints on NS radius. By requiring that a realistic EOS of NS matter must accommodate 
the NS maximal mass $M_{\rm max}\gtrsim 1.97~M_\odot$, Abbott {\it et al.} have obtained 
the radius of NS with $M\approx 1.4~M_\odot$ in the range $R_{1.4}\approx 11.9\pm 1.4$ 
km at the 90\% confidence level \cite{Abbott18}. This analysis has also given a 
constraint on the total pressure of NS matter at supra-saturation densities, namely, 
$P(2n_0)\approx 3.5^{+2.7}_{-1.7}\times 10^{34}$ dyn/cm$^2$ at the same 90\% confidence
level. One can see from NS properties given by the EOS of the spin-polarized 
$\beta$-stable np$e\mu$ matter shown in Table~\ref{t4} that the GW170817 constraints 
are fulfilled with the spin polarization of baryons $\Delta \lesssim 0.6$ for 
the $R_{1.4}$ radius and $\Delta\lesssim 1$ for the total pressure $P(2n_0)$ 
of NS matter, respectively. The NS maximal mass $M_{\rm max}$ obtained with the EOS 
given by the CDM3Y8 interaction increases from $1.98~~M_\odot$ to $2.16~M_\odot$ with 
the spin polarization $\Delta$ increasing from 0 to 1. This range of the $M_{\rm max}$ 
values covers well the observed NS masses $M\approx 1.908\pm 0.016~M_\odot$,  
$2.01\pm 0.04~M_\odot$, and $2.14\pm 0.09~M_\odot$ of the binary pulsars 
PSR J1614-2230 \cite{Demorest,Arzou18}, PSR J0348+0432 \cite{Sci340}, and PSR J0740+6620 
\cite{Cro19}, respectively. Note that the large NS mass $M\approx 2.14~M_\odot$ seems 
possible in the present mean-field scenario only when baryons are completely spin
polarized ($0.9\lesssim\Delta\lesssim 1$).
\begin{figure}[bht]\vspace{-2cm}
\includegraphics[angle=0,width=0.9\textwidth]{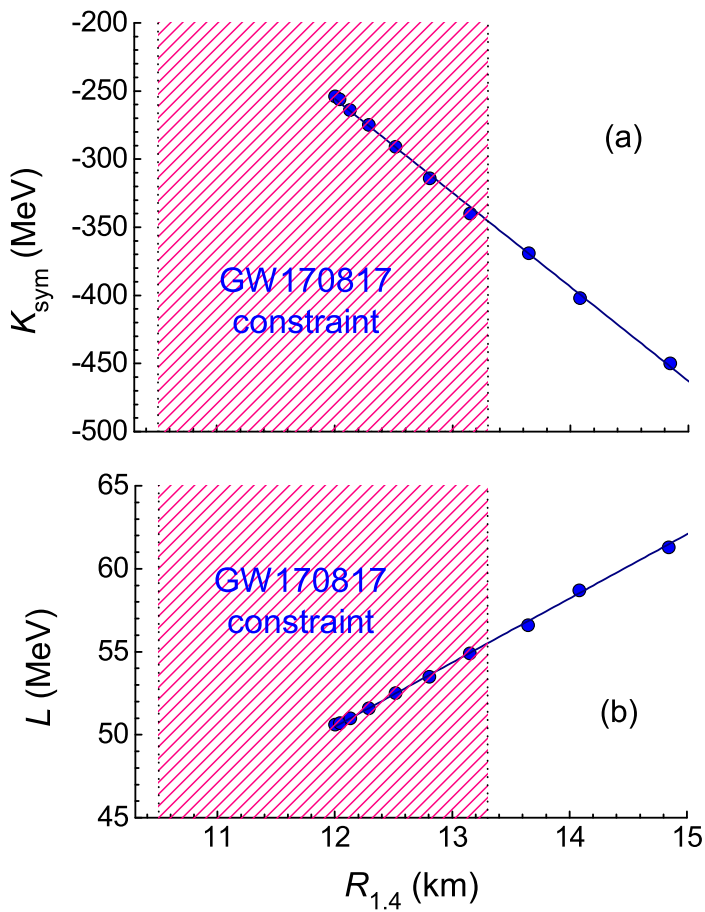}\vspace{-3cm}
\caption{The curvature $K_{\rm sym}$ and slope $L$ of the nuclear symmetry energy (\ref{eq8}) 
versus the radius $R_{1.4}$ of NS with $M\approx 1.4~M_\odot$ shown as solid circles in 
panels (a) and (b), respectively, are given by the EOS of the spin-polarized 
$\beta$-stable np$e\mu$ matter with $\Delta$ increasing from 0 to 0.9 (see also
Tables~\ref{t3} and \ref{t4}). The shaded areas enclose the range of the $R_{1.4}$ 
radius constrained by the tidal deformability of NS \cite{Abbott18}.} \label{f6} 
\end{figure} 
 
The constraint on the radius $R_{1.4}$ of NS with $M\approx 1.4~M_\odot$ deduced
from the multimessenger observation of GW170817 \cite{Abbott18} has now become 
an important reference for the mean-field studies or Bayesian analyses of the EOS 
of NS matter to narrow the uncertainty of the symmetry energy (\ref{eq8}) 
at high baryon densities \cite{Xie19,Tsang19}. For example, Tsang {\it et al.} 
 \cite{Tsang19} have shown a systematic correlation of the $J$, $L$, and $K_{\rm sym}$ 
values with the tidal deformability of NS, using about 200 different sets of Skyrme 
interaction in their mean-field study. From the GW170817 constraint on the radius 
$R_{1.4}$ and tidal deformability, some correlation between the symmetry energy 
at high baryon densities and $R_{1.4}$ radius can be inferred. By comparing the 
$J$, $L$, and $K_{\rm sym}$ values given by our HF calculation of the spin-polarized 
NM shown in Table~\ref{t3} and $R_{1.4}$ radii obtained at different $\Delta$ values 
shown in Table~\ref{t4}, we found that $J$, $L$, and $|K_{\rm sym}|$ are almost linearly 
correlated with the $R_{1.4}$ radius (see Fig.~\ref{f6}). In particular, with the 
increasing spin polarization of baryons, the larger the slope $L$ of the symmetry 
energy, the larger the corresponding $R_{1.4}$ radius. One can see in Fig.~\ref{f6} 
that the spin polarization of baryons is confined by the GW170817 constraint to 
the range $0\lesssim\Delta\lesssim 0.6$, where $50.6\lesssim L\lesssim 54.9$ MeV 
and $12\lesssim R_{1.4}\lesssim 13.1$ km. From the correlation between $L$ and 
$R_{1.4}$ shown in Fig.~\ref{f6}, it is not excluded that an EOS of the spin-unpolarized
NS matter with $L\gtrsim 65$ MeV would give a radius $R_{1.4} > 13.1$ km.

\begin{figure}\vspace{-1.5cm}\hspace*{0cm}
\includegraphics[angle=0,width=0.7\textwidth]{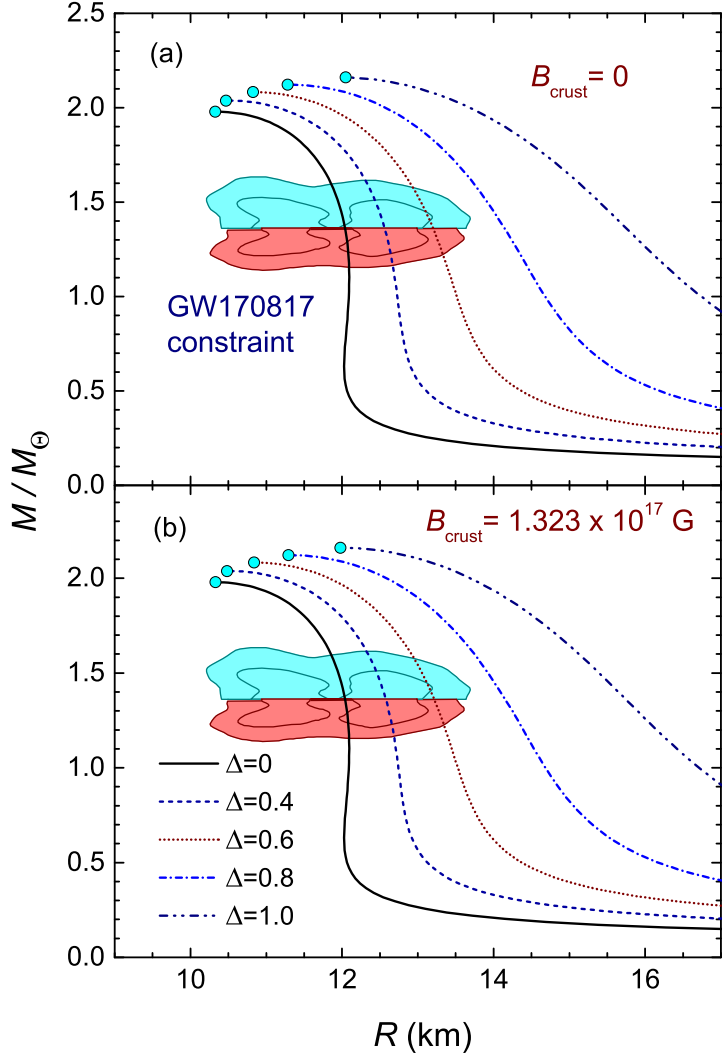}\vspace{-0.5cm}
 \caption{Correlation of the mass and radius of NS given by the EOS of the 
 $\beta$-stable spin-polarized np$e\mu$ matter ($\Delta=0\to 1$) obtained with 
 the CDM3Y8 interaction (\ref{eq2}). The results obtained with the unmagnetized 
 and magnetized NS crust given by the EDF theory \cite{Chamel19} are shown in 
 panels (a) and (b), respectively. The GW170817 constraint \cite{Abbott18} is enclosed 
 in the colored contours, and circles are the $M,R$ values at the maximum central 
 densities $n_c$.} \label{f7}
\end{figure} 
Our results obtained for the correlation of the mass and radius of NS given by the EOS 
of the $\beta$-stable spin-polarized np$e\mu$ matter with $\Delta$ increasing from 
0 to 1 are shown in Fig.~\ref{f7}. The GW170817 constraint for the radius $R_{1.4}$ 
of NS with $M\approx 1.4~M_\odot$ are plotted in Fig.~\ref{f7} as the shaded contours. 
One can see that all mass-radius curves with $\Delta\lesssim 0.6$ go well through 
the GW170817 contours. The upper bound of 13.3 km for $R_{1.4}$ \cite{Abbott18} 
is exceeded when more than 60\% of baryons are spin polarized. Combined with the
impact on the symmetry energy shown in Figs.~\ref{f3} and \ref{f6}, the upper limit
of $\Delta\approx 0.6$ shown in Fig.~\ref{f7} might well narrow the uncertainty 
of the symmetry energy $S$, its slope $L$, and curvature $K_{\rm sym}$ at 
supra-saturation baryon densities ($n_{\rm b}>n_0$). As shown in Table~\ref{t4} and
Fig.~\ref{f7}, the maximal gravitional mass $M_{\rm max}$ and radius $R_{\rm max}$ 
of NS in the hydrostatic equilibrium can be strongly affected by the spin polarization 
of baryons. We note that all the calculated $M_{\rm max}$ values are above the 
lower limit of $1.97~M_\odot$ that was imposed on the GW170817 constraint for the 
$R_{1.4}$ radius by Abbott {\it et al.} \cite{Abbott18}. The results shown in the 
upper- and lower panels of Fig.~\ref{f7} confirm that the effect of the magnetization 
of the NS crust \cite{Chamel19} on the NS mass and radius is negligible. It remains
uncertain if this conclusion still holds when the spin polarization of free baryons 
as well as the spin-unsaturated nuclei in the Wigner-Seitz cell in the NS crust 
is treated explicitly.

In conclusion, the present mean-field study shows that the spin polarization 
of baryons affects strongly the EOS of the NS matter via the spin- and spin-isospin 
dependent channels of the in-medium interaction between baryons. In particular, 
the symmetry energy of the spin-polarized  NM (with $\Delta\neq 0$) was found 
to be much stiffer at high baryon densities compared with that of the spin-saturated 
NM (with $\Delta=0$), and this can affect significantly the hydrostatic configuration
of NS. Based on the GW170817 constraint for the $R_{1.4}$ radius \cite{Abbott18}, 
baryons in the two merging neutron stars could be partially spin-polarized 
with $0\lesssim\Delta\lesssim 0.6$ (see Fig.~\ref{f7}). We found, however, that 
the GW170817 constraint excludes the full spin polarization of baryons in NS 
matter ($\Delta=1$), and this results agrees well with a recent 
conclusion by Tews and Schwenk \cite{Tews20}.

\subsection*{Possible effects by the density dependence of the spin polarization}
We note that the above results have been obtained with the spin polarization 
of baryons assumed to be independent of the baryon density $n_{\rm b}$. However, 
the magnetic-field distribution in the NS matter is quite complex \cite{Fuji14}, and 
the spin polarization of baryons in a magnetar is expected to be gradually weakened 
with the increasing baryon density. In particular, with the magnetic-field intensity 
diminishing to zero in the NS center \cite{Fuji14}, the spin polarization 
of baryons is also expected to decrease to $\Delta\approx 0$ in the central region 
of magnetar. Although it is beyond the scope of the present mean-field approach
to properly calculate the density profile $\Delta(n_{\rm b})$ of the spin polarization 
of baryons in magnetar, we try to explore this effect by assuming three simple scenarios 
(A,B, and C) for the density dependence of $\Delta$ based on the magnetic-field 
distribution in magnetar obtained by Fujisawa and Kisaka using the Green's function 
relaxation method (see lower panel of Fig.~3 in Ref.~\cite{Fuji14}). 

A) The magnetic field is strongly localized in the surface region of magnetar, 
around the crust-core transition, and quickly decreases to $B\approx 0$ at the 
baryon density $n_{\rm b}\approx 0.18$ fm$^{-3}$. We consider explicitly the spin 
polarization of baryons $\Delta=0.6,\ 0.8$, and 1.0, which are assumed 
to gradually weaken to $\Delta\approx 0$ at this same density.

B) The distribution of the magnetic-field strength is broader and covers both 
the crust and outer core of magnetar, so that $\Delta$ decreases smoothly 
to zero at a larger baryon density $n_{\rm b}\approx 0.35$ fm$^{-3}$.

C) The magnetic-field strength is spreading to even higher baryon densities 
and decreasing to $B\approx 0$ at $n_{\rm b}\approx 0.5$ fm$^{-3}$. Three 
considered strengths of the spin polarization of baryons are gradually weakening 
to $\Delta\approx 0$ at this same density. 

\begin{figure}\vspace{-0.5cm}\hspace*{-0.5cm}
\includegraphics[angle=0,width=1.1\textwidth]{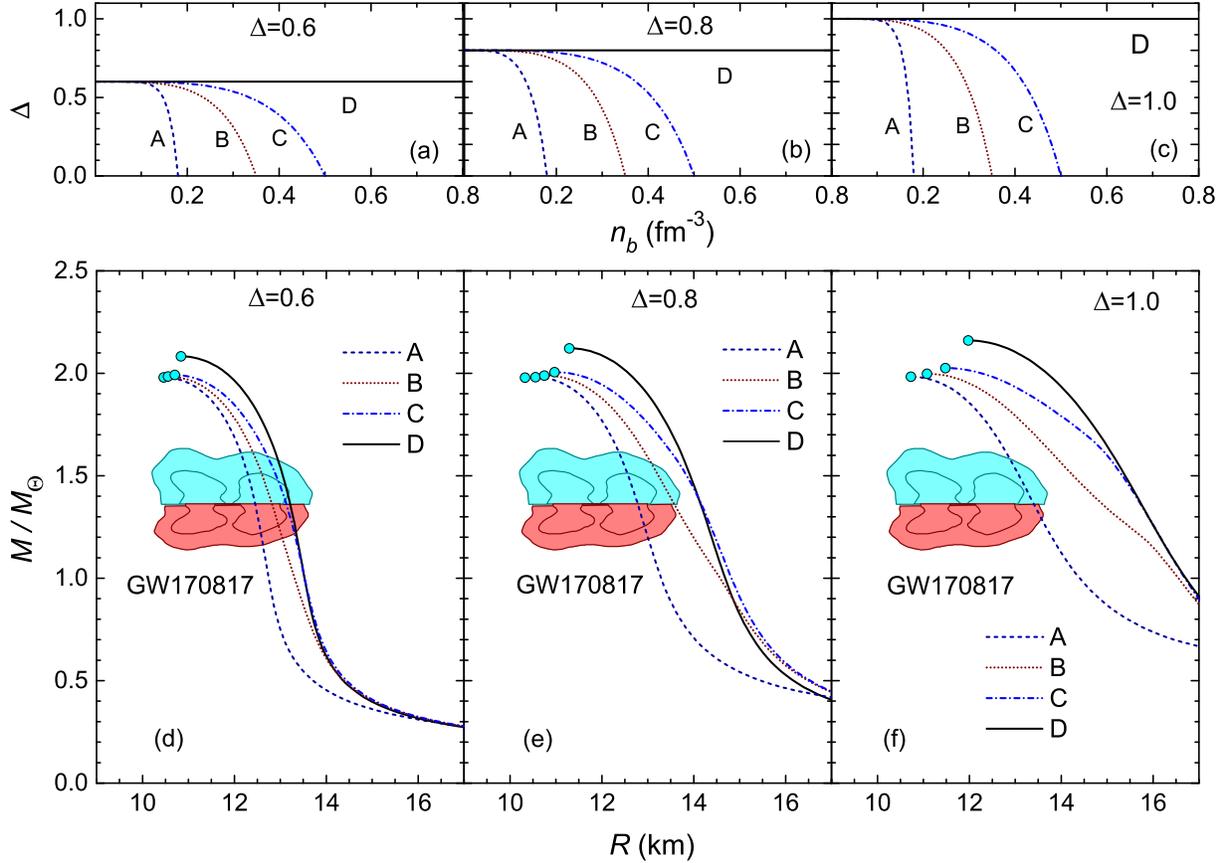}\vspace{-0cm}
 \caption{Scenarios A,B,C, and D for the density-dependent spin polarization 
 of baryons with $\Delta$ starting from 0.6, 0.8, and 1 are shown in panels (a), 
(b), and (c), respectively. The corresponding mass and radius of magnetar given by the 
EOS of the $\beta$-stable  spin-polarized np$e\mu$ matter are shown in panels (d), (e), 
and (f), respectively. The colored countours enclose the region allowed by the GW170817 
constraint \cite{Abbott18}.} \label{f8}
\end{figure}

Thus, the magnetic field in these three scenarios is completely depleted ($B=0$) in 
the central region of magnetar where the baryon density $n_{\rm b}$ approaches 
$0.6\sim 1$ fm$^{-3}$. 
The suggested density-dependent profiles of $\Delta$ are shown together with the 
uniform (density independent) spin polarization of baryons (scenarios D) in the upper 
panel of Fig.~\ref{f8}. The corresponding TOV results for the mass and radius 
of magnetar given by the density-dependent spin polarization of baryons are shown 
in the lower panel of Fig.~\ref{f8}, where the GW170817 constraint for the $R_{1.4}$  
radius are plotted as the shaded contours. One can see that when the magnetic-field 
strength is localized narrowly in the crust-core transition (scenario A), the mass 
and radius of magnetar obtained with $\Delta\lesssim 1$ are within the boundaries 
of the GW170817 constraint. However, with the magnetic-field strength spreading 
more into to the outer core (scenarios A and B) the full spin polarization of baryons 
($\Delta=1$) is ruled out, and only a partial spin polarization of baryons with 
$\Delta\lesssim 0.8$ is possible. It is noteworthy that the EOS of partially 
spin-polarized NS matter with $0\lesssim\Delta\lesssim 0.6$ gives the mass and 
radius of magnetar well within the GW170817 boundaries in all three scenarios 
(see left panel of Fig.~\ref{f8}). 

In general, when the electromagnetic interaction between the magnetic field and 
NS matter is taken into account explicitly, the $B$-dependent contribution to the 
total energy density of the spin-polarized NS matter (\ref{eq9}) is not negligible
\cite{Agui14,Isayev12}, and the discussed effects of the spin polarization 
of baryons to the hydrostatic NS configuration might become even more significant.  
Given the results of previous studies \cite{Latt07,Bro00,Agui14,Isayev12,Fuji14} 
on the effects of magnetic field, the total impact on the EOS of magnetar 
matter by magnetic field might become essential at the moderate field strength, 
as found in the magnetar scenario by Metzger {\it et al.} \cite{Metzger18} for 
the blue kilonova ejecta of GW170817.      

\section*{Summary}  
The nonrelativistic HF approach \cite{Loan11} has been extended to study the 
spin-polarized NM using the new density-dependent CDM3Y8 version of the M3Y 
interaction \cite{Anan83}, with its spin- and spin-isospin dependence adjusted 
to obtain the HF results close to those of the BHF calculation of the spin-polarized 
neutron matter \cite{Vidana16}. Like for the nuclear (isospin) symmetry energy $S$, 
the parabolic approximation was found to be valid also for the spin-symmetry energy $W$,
so that the (repulsive) contribution to the HF energy density of NM from the spin 
polarization of baryons is directly proportional to $\Delta^2$, and the EOS of 
NM becomes stiffer with the increasing spin polarization of baryons.   

The $\Delta$ dependence of the symmetry coefficient $J$, the slope $L$, and 
curvature $K_{\rm sym}$ of the nuclear symmetry energy $S$ has been investigated, 
and we found that the empirical ranges adopted for these quantities \cite{Zhang18} 
include results of the present HF calculation with the spin polarization of baryons 
up to $\Delta\approx 0.8$. With the increasing $\Delta$, the $J$, $L$, and $K_{\rm sym}$ 
values were found to correlate linearly with the radius $R_{1.4}$ of a NS with mass
$M\approx 1.4~M_\odot$.    

The total HF energy density of the $\beta$-stable np$e\mu$ matter has been obtained
at different spin polarization of baryons, and the proton fraction $x_{\rm p}$ 
was found to increase strongly with increasing $\Delta$, which in turn leads 
readily to a larger probability of the direct Urca process in the cooling of the magnetar. 

The stiffening of the symmetry energy of the $\beta$-stable spin-polarized  
np$e\mu$ matter at high baryon densities has been shown to affect significantly 
the hydrostatic NS configuration. By subjecting the mass and radius of a NS obtained 
at different spin polarizations of baryons to the GW170817 constraint on 
the $R_{1.4}$ radius \cite{Abbott18}, we found that up to 60\% of baryons 
might have their spins polarized during the NS merger. The same conclusion 
can be made when $\Delta$ is assumed to be density dependent, and the spin 
polarization of baryons is gradually decreasing from the surface of magnetar 
to zero at $n_{\rm b}\lesssim~3n_0$. These results support the magnetar 
origin of the blue kilonova ejecta of GW170817 suggested 
by Metzger {\it et al.} \cite{Metzger18}.   

\section*{Acknowledgement}
The present research was supported, in part, by the National Foundation for
Science and Technology Development of Vietnam (NAFOSTED Project 
No. 103.04-2017.317). We also thank Isaac Vida\~na for his helpful communication 
on the BHF results of the spin-polarized neutron matter \cite{Vidana16}, and 
Nicolas Chamel for providing us with the EOS of the (unmagnetized and magnetized) 
NS crust \cite{Chamel18,Chamel19} in the tabulated form.

\bibliographystyle{apsrev4-1}
\bibliography{pTan3R}

\begin{thebibliography}{43}%
\makeatletter
\providecommand \@ifxundefined [1]{%
 \@ifx{#1\undefined}
}%
\providecommand \@ifnum [1]{%
 \ifnum #1\expandafter \@firstoftwo
 \else \expandafter \@secondoftwo
 \fi
}%
\providecommand \@ifx [1]{%
 \ifx #1\expandafter \@firstoftwo
 \else \expandafter \@secondoftwo
 \fi
}%
\providecommand \natexlab [1]{#1}%
\providecommand \enquote  [1]{``#1''}%
\providecommand \bibnamefont  [1]{#1}%
\providecommand \bibfnamefont [1]{#1}%
\providecommand \citenamefont [1]{#1}%
\providecommand \href@noop [0]{\@secondoftwo}%
\providecommand \href [0]{\begingroup \@sanitize@url \@href}%
\providecommand \@href[1]{\@@startlink{#1}\@@href}%
\providecommand \@@href[1]{\endgroup#1\@@endlink}%
\providecommand \@sanitize@url [0]{\catcode `\\12\catcode `\$12\catcode
  `\&12\catcode `\#12\catcode `\^12\catcode `\_12\catcode `\%12\relax}%
\providecommand \@@startlink[1]{}%
\providecommand \@@endlink[0]{}%
\providecommand \url  [0]{\begingroup\@sanitize@url \@url }%
\providecommand \@url [1]{\endgroup\@href {#1}{\urlprefix }}%
\providecommand \urlprefix  [0]{URL }%
\providecommand \Eprint [0]{\href }%
\providecommand \doibase [0]{http://dx.doi.org/}%
\providecommand \selectlanguage [0]{\@gobble}%
\providecommand \bibinfo  [0]{\@secondoftwo}%
\providecommand \bibfield  [0]{\@secondoftwo}%
\providecommand \translation [1]{[#1]}%
\providecommand \BibitemOpen [0]{}%
\providecommand \bibitemStop [0]{}%
\providecommand \bibitemNoStop [0]{.\EOS\space}%
\providecommand \EOS [0]{\spacefactor3000\relax}%
\providecommand \BibitemShut  [1]{\csname bibitem#1\endcsname}%
\let\auto@bib@innerbib\@empty
\bibitem [{\citenamefont {Lattimer}\ and\ \citenamefont
  {Prakash}(2007)}]{Latt07}%
  \BibitemOpen
  \bibfield  {author} {\bibinfo {author} {\bibfnamefont {J.~M.}\ \bibnamefont
  {Lattimer}}\ and\ \bibinfo {author} {\bibfnamefont {M.}~\bibnamefont
  {Prakash}},\ }\href {\doibase https://doi.org/10.1016/j.physrep.2007.02.003}
  {\bibfield  {journal} {\bibinfo  {journal} {Phys. Rep.}\ }\textbf {\bibinfo
  {volume} {442}},\ \bibinfo {pages} {109 } (\bibinfo {year}
  {2007})}\BibitemShut {NoStop}%
\bibitem [{\citenamefont {Broderick}\ \emph {et~al.}(2000)\citenamefont
  {Broderick}, \citenamefont {Prakash},\ and\ \citenamefont
  {Lattimer}}]{Bro00}%
  \BibitemOpen
  \bibfield  {author} {\bibinfo {author} {\bibfnamefont {A.}~\bibnamefont
  {Broderick}}, \bibinfo {author} {\bibfnamefont {M.}~\bibnamefont {Prakash}},
  \ and\ \bibinfo {author} {\bibfnamefont {J.~M.}\ \bibnamefont {Lattimer}},\
  }\href {\doibase 10.1086/309010} {\bibfield  {journal} {\bibinfo  {journal}
  {Astrophys. J.}\ }\textbf {\bibinfo {volume} {537}},\ \bibinfo {pages} {351}
  (\bibinfo {year} {2000})}\BibitemShut {NoStop}%
\bibitem [{\citenamefont {Dexheimer}\ \emph {et~al.}(2017)\citenamefont
  {Dexheimer}, \citenamefont {Franzon}, \citenamefont {Gomes}, \citenamefont
  {Farias}, \citenamefont {Avancini},\ and\ \citenamefont {Schramm}}]{Dex17}%
  \BibitemOpen
  \bibfield  {author} {\bibinfo {author} {\bibfnamefont {V.}~\bibnamefont
  {Dexheimer}}, \bibinfo {author} {\bibfnamefont {B.}~\bibnamefont {Franzon}},
  \bibinfo {author} {\bibfnamefont {R.}~\bibnamefont {Gomes}}, \bibinfo
  {author} {\bibfnamefont {R.}~\bibnamefont {Farias}}, \bibinfo {author}
  {\bibfnamefont {S.}~\bibnamefont {Avancini}}, \ and\ \bibinfo {author}
  {\bibfnamefont {S.}~\bibnamefont {Schramm}},\ }\href {\doibase
  https://doi.org/10.1016/j.physletb.2017.09.008} {\bibfield  {journal}
  {\bibinfo  {journal} {Phys. Lett. B}\ }\textbf {\bibinfo {volume} {773}},\
  \bibinfo {pages} {487 } (\bibinfo {year} {2017})}\BibitemShut {NoStop}%
\bibitem [{\citenamefont {Abbott~\textit{et
  al.}}(2017{\natexlab{a}})}]{Abbott17}%
  \BibitemOpen
  \bibfield  {author} {\bibinfo {author} {\bibfnamefont {B.~P.}\ \bibnamefont
  {Abbott~\textit{et al.}}},\ }\href {\doibase 10.1103/PhysRevLett.119.161101}
  {\bibfield  {journal} {\bibinfo  {journal} {Phys. Rev. Lett.}\ }\textbf
  {\bibinfo {volume} {119}},\ \bibinfo {pages} {161101} (\bibinfo {year}
  {2017}{\natexlab{a}})}\BibitemShut {NoStop}%
\bibitem [{\citenamefont {Abbott~\textit{et al.}}(2018)}]{Abbott18}%
  \BibitemOpen
  \bibfield  {author} {\bibinfo {author} {\bibfnamefont {B.~P.}\ \bibnamefont
  {Abbott~\textit{et al.}}},\ }\href {\doibase 10.1103/PhysRevLett.121.161101}
  {\bibfield  {journal} {\bibinfo  {journal} {Phys. Rev. Lett.}\ }\textbf
  {\bibinfo {volume} {121}},\ \bibinfo {pages} {161101} (\bibinfo {year}
  {2018})}\BibitemShut {NoStop}%
\bibitem [{\citenamefont {Abbott~\textit{et
  al.}}(2017{\natexlab{b}})}]{Abbott17c}%
  \BibitemOpen
  \bibfield  {author} {\bibinfo {author} {\bibfnamefont {B.~P.}\ \bibnamefont
  {Abbott~\textit{et al.}}},\ }\href {\doibase 10.3847/2041-8213/aa91c9}
  {\bibfield  {journal} {\bibinfo  {journal} {Astrophys. J. Lett.}\ }\textbf
  {\bibinfo {volume} {848}},\ \bibinfo {pages} {L12} (\bibinfo {year}
  {2017}{\natexlab{b}})}\BibitemShut {NoStop}%
\bibitem [{\citenamefont {Evans~\textit{et al.}}(2017)}]{Evan17}%
  \BibitemOpen
  \bibfield  {author} {\bibinfo {author} {\bibfnamefont {P.~A.}\ \bibnamefont
  {Evans~\textit{et al.}}},\ }\href {\doibase 10.1126/science.aap9580}
  {\bibfield  {journal} {\bibinfo  {journal} {Science}\ }\textbf {\bibinfo
  {volume} {358}},\ \bibinfo {pages} {1565} (\bibinfo {year}
  {2017})}\BibitemShut {NoStop}%
\bibitem [{\citenamefont {Metzger}\ \emph {et~al.}(2018)\citenamefont
  {Metzger}, \citenamefont {Thompson},\ and\ \citenamefont
  {Quataert}}]{Metzger18}%
  \BibitemOpen
  \bibfield  {author} {\bibinfo {author} {\bibfnamefont {B.~D.}\ \bibnamefont
  {Metzger}}, \bibinfo {author} {\bibfnamefont {T.~A.}\ \bibnamefont
  {Thompson}}, \ and\ \bibinfo {author} {\bibfnamefont {E.}~\bibnamefont
  {Quataert}},\ }\href {\doibase 10.3847/1538-4357/aab095} {\bibfield
  {journal} {\bibinfo  {journal} {Astrophys. J.}\ }\textbf {\bibinfo {volume}
  {856}},\ \bibinfo {pages} {101} (\bibinfo {year} {2018})}\BibitemShut
  {NoStop}%
\bibitem [{\citenamefont {Fujisawa}\ and\ \citenamefont
  {Kisaka}(2014)}]{Fuji14}%
  \BibitemOpen
  \bibfield  {author} {\bibinfo {author} {\bibfnamefont {K.}~\bibnamefont
  {Fujisawa}}\ and\ \bibinfo {author} {\bibfnamefont {S.}~\bibnamefont
  {Kisaka}},\ }\href {\doibase 10.1093/mnras/stu1911} {\bibfield  {journal}
  {\bibinfo  {journal} {Mon. Not. Roy. Astron. Soc.}\ }\textbf {\bibinfo
  {volume} {445}},\ \bibinfo {pages} {2777} (\bibinfo {year}
  {2014})}\BibitemShut {NoStop}%
\bibitem [{\citenamefont {Vida\~na}\ and\ \citenamefont
  {Bombaci}(2002)}]{Vidana02}%
  \BibitemOpen
  \bibfield  {author} {\bibinfo {author} {\bibfnamefont {I.}~\bibnamefont
  {Vida\~na}}\ and\ \bibinfo {author} {\bibfnamefont {I.}~\bibnamefont
  {Bombaci}},\ }\href {\doibase 10.1103/PhysRevC.66.045801} {\bibfield
  {journal} {\bibinfo  {journal} {Phys. Rev. C}\ }\textbf {\bibinfo {volume}
  {66}},\ \bibinfo {pages} {045801} (\bibinfo {year} {2002})}\BibitemShut
  {NoStop}%
\bibitem [{\citenamefont {Vida\~na}\ \emph {et~al.}(2016)\citenamefont
  {Vida\~na}, \citenamefont {Polls},\ and\ \citenamefont {Durant}}]{Vidana16}%
  \BibitemOpen
  \bibfield  {author} {\bibinfo {author} {\bibfnamefont {I.}~\bibnamefont
  {Vida\~na}}, \bibinfo {author} {\bibfnamefont {A.}~\bibnamefont {Polls}}, \
  and\ \bibinfo {author} {\bibfnamefont {V.}~\bibnamefont {Durant}},\ }\href
  {\doibase 10.1103/PhysRevC.94.054006} {\bibfield  {journal} {\bibinfo
  {journal} {Phys. Rev. C}\ }\textbf {\bibinfo {volume} {94}},\ \bibinfo
  {pages} {054006} (\bibinfo {year} {2016})}\BibitemShut {NoStop}%
\bibitem [{\citenamefont {Aguirre}\ \emph {et~al.}(2014)\citenamefont
  {Aguirre}, \citenamefont {Bauer},\ and\ \citenamefont {Vidana}}]{Agui14}%
  \BibitemOpen
  \bibfield  {author} {\bibinfo {author} {\bibfnamefont {R.}~\bibnamefont
  {Aguirre}}, \bibinfo {author} {\bibfnamefont {E.}~\bibnamefont {Bauer}}, \
  and\ \bibinfo {author} {\bibfnamefont {I.}~\bibnamefont {Vidana}},\ }\href
  {\doibase 10.1103/PhysRevC.89.035809} {\bibfield  {journal} {\bibinfo
  {journal} {Phys. Rev. C}\ }\textbf {\bibinfo {volume} {89}},\ \bibinfo
  {pages} {035809} (\bibinfo {year} {2014})}\BibitemShut {NoStop}%
\bibitem [{\citenamefont {Isayev}\ and\ \citenamefont {Yang}(2012)}]{Isayev12}%
  \BibitemOpen
  \bibfield  {author} {\bibinfo {author} {\bibfnamefont {A.}~\bibnamefont
  {Isayev}}\ and\ \bibinfo {author} {\bibfnamefont {J.}~\bibnamefont {Yang}},\
  }\href {\doibase https://doi.org/10.1016/j.physletb.2011.12.003} {\bibfield
  {journal} {\bibinfo  {journal} {Phys. Lett. B}\ }\textbf {\bibinfo {volume}
  {707}},\ \bibinfo {pages} {163 } (\bibinfo {year} {2012})}\BibitemShut
  {NoStop}%
\bibitem [{\citenamefont {Tews}\ and\ \citenamefont {Schwenk}(2020)}]{Tews20}%
  \BibitemOpen
  \bibfield  {author} {\bibinfo {author} {\bibfnamefont {I.}~\bibnamefont
  {Tews}}\ and\ \bibinfo {author} {\bibfnamefont {A.}~\bibnamefont {Schwenk}},\
  }\href {\doibase 10.3847/1538-4357/ab7232} {\bibfield  {journal} {\bibinfo
  {journal} {Astrophys. J.}\ }\textbf {\bibinfo {volume} {892}},\ \bibinfo
  {pages} {14} (\bibinfo {year} {2020})}\BibitemShut {NoStop}%
\bibitem [{\citenamefont {Loan}\ \emph {et~al.}(2011)\citenamefont {Loan},
  \citenamefont {Tan}, \citenamefont {Khoa},\ and\ \citenamefont
  {Margueron}}]{Loan11}%
  \BibitemOpen
  \bibfield  {author} {\bibinfo {author} {\bibfnamefont {D.~T.}\ \bibnamefont
  {Loan}}, \bibinfo {author} {\bibfnamefont {N.~H.}\ \bibnamefont {Tan}},
  \bibinfo {author} {\bibfnamefont {D.~T.}\ \bibnamefont {Khoa}}, \ and\
  \bibinfo {author} {\bibfnamefont {J.}~\bibnamefont {Margueron}},\ }\href
  {\doibase 10.1103/PhysRevC.83.065809} {\bibfield  {journal} {\bibinfo
  {journal} {Phys. Rev. C}\ }\textbf {\bibinfo {volume} {83}},\ \bibinfo
  {pages} {065809} (\bibinfo {year} {2011})}\BibitemShut {NoStop}%
\bibitem [{\citenamefont {Tan}\ \emph {et~al.}(2016)\citenamefont {Tan},
  \citenamefont {Loan}, \citenamefont {Khoa},\ and\ \citenamefont
  {Margueron}}]{Tan16}%
  \BibitemOpen
  \bibfield  {author} {\bibinfo {author} {\bibfnamefont {N.~H.}\ \bibnamefont
  {Tan}}, \bibinfo {author} {\bibfnamefont {D.~T.}\ \bibnamefont {Loan}},
  \bibinfo {author} {\bibfnamefont {D.~T.}\ \bibnamefont {Khoa}}, \ and\
  \bibinfo {author} {\bibfnamefont {J.}~\bibnamefont {Margueron}},\ }\href
  {\doibase 10.1103/PhysRevC.93.035806} {\bibfield  {journal} {\bibinfo
  {journal} {Phys. Rev. C}\ }\textbf {\bibinfo {volume} {93}},\ \bibinfo
  {pages} {035806} (\bibinfo {year} {2016})}\BibitemShut {NoStop}%
\bibitem [{\citenamefont {Khoa}\ \emph {et~al.}(1997)\citenamefont {Khoa},
  \citenamefont {Satchler},\ and\ \citenamefont {von Oertzen}}]{Kho97}%
  \BibitemOpen
  \bibfield  {author} {\bibinfo {author} {\bibfnamefont {D.~T.}\ \bibnamefont
  {Khoa}}, \bibinfo {author} {\bibfnamefont {G.~R.}\ \bibnamefont {Satchler}},
  \ and\ \bibinfo {author} {\bibfnamefont {W.}~\bibnamefont {von Oertzen}},\
  }\href {\doibase 10.1103/PhysRevC.56.954} {\bibfield  {journal} {\bibinfo
  {journal} {Phys. Rev. C}\ }\textbf {\bibinfo {volume} {56}},\ \bibinfo
  {pages} {954} (\bibinfo {year} {1997})}\BibitemShut {NoStop}%
\bibitem [{\citenamefont {Khoa}\ and\ \citenamefont {Satchler}(2000)}]{Kho00}%
  \BibitemOpen
  \bibfield  {author} {\bibinfo {author} {\bibfnamefont {D.~T.}\ \bibnamefont
  {Khoa}}\ and\ \bibinfo {author} {\bibfnamefont {G.~R.}\ \bibnamefont
  {Satchler}},\ }\href {\doibase https://doi.org/10.1016/S0375-9474(99)00680-6}
  {\bibfield  {journal} {\bibinfo  {journal} {Nucl. Phys. A}\ }\textbf
  {\bibinfo {volume} {668}},\ \bibinfo {pages} {3 } (\bibinfo {year}
  {2000})}\BibitemShut {NoStop}%
\bibitem [{\citenamefont {Anantaraman}\ \emph {et~al.}(1983)\citenamefont
  {Anantaraman}, \citenamefont {Toki},\ and\ \citenamefont {Bertsch}}]{Anan83}%
  \BibitemOpen
  \bibfield  {author} {\bibinfo {author} {\bibfnamefont {N.}~\bibnamefont
  {Anantaraman}}, \bibinfo {author} {\bibfnamefont {H.}~\bibnamefont {Toki}}, \
  and\ \bibinfo {author} {\bibfnamefont {G.}~\bibnamefont {Bertsch}},\ }\href
  {\doibase https://doi.org/10.1016/0375-9474(83)90487-6} {\bibfield  {journal}
  {\bibinfo  {journal} {Nucl. Phys. A}\ }\textbf {\bibinfo {volume} {398}},\
  \bibinfo {pages} {269 } (\bibinfo {year} {1983})}\BibitemShut {NoStop}%
\bibitem [{\citenamefont {Khoa}\ \emph {et~al.}(1996)\citenamefont {Khoa},
  \citenamefont {von Oertzen},\ and\ \citenamefont {Ogloblin}}]{Kho96}%
  \BibitemOpen
  \bibfield  {author} {\bibinfo {author} {\bibfnamefont {D.~T.}\ \bibnamefont
  {Khoa}}, \bibinfo {author} {\bibfnamefont {W.}~\bibnamefont {von Oertzen}}, \
  and\ \bibinfo {author} {\bibfnamefont {A.}~\bibnamefont {Ogloblin}},\ }\href
  {\doibase https://doi.org/10.1016/0375-9474(96)00091-7} {\bibfield  {journal}
  {\bibinfo  {journal} {Nucl. Phys. A}\ }\textbf {\bibinfo {volume} {602}},\
  \bibinfo {pages} {98 } (\bibinfo {year} {1996})}\BibitemShut {NoStop}%
\bibitem [{\citenamefont {Akmal}\ \emph {et~al.}(1998)\citenamefont {Akmal},
  \citenamefont {Pandharipande},\ and\ \citenamefont {Ravenhall}}]{APR}%
  \BibitemOpen
  \bibfield  {author} {\bibinfo {author} {\bibfnamefont {A.}~\bibnamefont
  {Akmal}}, \bibinfo {author} {\bibfnamefont {V.~R.}\ \bibnamefont
  {Pandharipande}}, \ and\ \bibinfo {author} {\bibfnamefont {D.~G.}\
  \bibnamefont {Ravenhall}},\ }\href {\doibase 10.1103/PhysRevC.58.1804}
  {\bibfield  {journal} {\bibinfo  {journal} {Phys. Rev. C}\ }\textbf {\bibinfo
  {volume} {58}},\ \bibinfo {pages} {1804} (\bibinfo {year}
  {1998})}\BibitemShut {NoStop}%
\bibitem [{\citenamefont {Gandolfi}\ \emph {et~al.}(2010)\citenamefont
  {Gandolfi}, \citenamefont {Illarionov}, \citenamefont {Fantoni},
  \citenamefont {Miller}, \citenamefont {Pederiva},\ and\ \citenamefont
  {Schmidt}}]{MMC}%
  \BibitemOpen
  \bibfield  {author} {\bibinfo {author} {\bibfnamefont {S.}~\bibnamefont
  {Gandolfi}}, \bibinfo {author} {\bibfnamefont {A.~Y.}\ \bibnamefont
  {Illarionov}}, \bibinfo {author} {\bibfnamefont {S.}~\bibnamefont {Fantoni}},
  \bibinfo {author} {\bibfnamefont {J.~C.}\ \bibnamefont {Miller}}, \bibinfo
  {author} {\bibfnamefont {F.}~\bibnamefont {Pederiva}}, \ and\ \bibinfo
  {author} {\bibfnamefont {K.~E.}\ \bibnamefont {Schmidt}},\ }\href@noop {}
  {\bibfield  {journal} {\bibinfo  {journal} {Mon. Not. Roy. Astron. Soc.}\
  }\textbf {\bibinfo {volume} {404}},\ \bibinfo {pages} {L35} (\bibinfo {year}
  {2010})}\BibitemShut {NoStop}%
\bibitem [{\citenamefont {Khoa}\ \emph
  {et~al.}(2007{\natexlab{a}})\citenamefont {Khoa}, \citenamefont {Than},\ and\
  \citenamefont {Cuong}}]{Kho07}%
  \BibitemOpen
  \bibfield  {author} {\bibinfo {author} {\bibfnamefont {D.~T.}\ \bibnamefont
  {Khoa}}, \bibinfo {author} {\bibfnamefont {H.~S.}\ \bibnamefont {Than}}, \
  and\ \bibinfo {author} {\bibfnamefont {D.~C.}\ \bibnamefont {Cuong}},\ }\href
  {\doibase 10.1103/PhysRevC.76.014603} {\bibfield  {journal} {\bibinfo
  {journal} {Phys. Rev. C}\ }\textbf {\bibinfo {volume} {76}},\ \bibinfo
  {pages} {014603} (\bibinfo {year} {2007}{\natexlab{a}})}\BibitemShut
  {NoStop}%
\bibitem [{\citenamefont {Khoa}\ \emph {et~al.}(2014)\citenamefont {Khoa},
  \citenamefont {Loc},\ and\ \citenamefont {Thang}}]{Kho14}%
  \BibitemOpen
  \bibfield  {author} {\bibinfo {author} {\bibfnamefont {D.~T.}\ \bibnamefont
  {Khoa}}, \bibinfo {author} {\bibfnamefont {B.~M.}\ \bibnamefont {Loc}}, \
  and\ \bibinfo {author} {\bibfnamefont {D.~N.}\ \bibnamefont {Thang}},\ }\href
  {\doibase https://doi.org/10.1140/epja/i2014-14034-9} {\bibfield  {journal}
  {\bibinfo  {journal} {Eur. Phys. J. A}\ }\textbf {\bibinfo {volume} {50}},\
  \bibinfo {pages} {34} (\bibinfo {year} {2014})}\BibitemShut {NoStop}%
\bibitem [{\citenamefont {Tsang}\ \emph {et~al.}(2011)\citenamefont {Tsang},
  \citenamefont {Chajecki}, \citenamefont {Coupland}, \citenamefont
  {Danielewicz}, \citenamefont {Famiano}, \citenamefont {Hodges}, \citenamefont
  {Kilburn}, \citenamefont {Lu}, \citenamefont {Lynch}, \citenamefont
  {Winkelbauer}, \citenamefont {Youngs},\ and\ \citenamefont
  {Zhang}}]{Tsang11}%
  \BibitemOpen
  \bibfield  {author} {\bibinfo {author} {\bibfnamefont {M.~B.}\ \bibnamefont
  {Tsang}}, \bibinfo {author} {\bibfnamefont {Z.}~\bibnamefont {Chajecki}},
  \bibinfo {author} {\bibfnamefont {D.}~\bibnamefont {Coupland}}, \bibinfo
  {author} {\bibfnamefont {P.}~\bibnamefont {Danielewicz}}, \bibinfo {author}
  {\bibfnamefont {F.}~\bibnamefont {Famiano}}, \bibinfo {author} {\bibfnamefont
  {R.}~\bibnamefont {Hodges}}, \bibinfo {author} {\bibfnamefont
  {M.}~\bibnamefont {Kilburn}}, \bibinfo {author} {\bibfnamefont
  {F.}~\bibnamefont {Lu}}, \bibinfo {author} {\bibfnamefont {W.~G.}\
  \bibnamefont {Lynch}}, \bibinfo {author} {\bibfnamefont {J.}~\bibnamefont
  {Winkelbauer}}, \bibinfo {author} {\bibfnamefont {M.}~\bibnamefont {Youngs}},
  \ and\ \bibinfo {author} {\bibfnamefont {Y.~X.}\ \bibnamefont {Zhang}},\
  }\href {\doibase https://doi.org/10.1016/j.ppnp.2011.01.041} {\bibfield
  {journal} {\bibinfo  {journal} {Prog. Part. Nucl. Phys.}\ }\textbf {\bibinfo
  {volume} {66}},\ \bibinfo {pages} {400 } (\bibinfo {year}
  {2011})}\BibitemShut {NoStop}%
\bibitem [{\citenamefont {Ono}\ \emph {et~al.}(2003)\citenamefont {Ono},
  \citenamefont {Danielewicz}, \citenamefont {Friedman}, \citenamefont
  {Lynch},\ and\ \citenamefont {Tsang}}]{Ono03}%
  \BibitemOpen
  \bibfield  {author} {\bibinfo {author} {\bibfnamefont {A.}~\bibnamefont
  {Ono}}, \bibinfo {author} {\bibfnamefont {P.}~\bibnamefont {Danielewicz}},
  \bibinfo {author} {\bibfnamefont {W.~A.}\ \bibnamefont {Friedman}}, \bibinfo
  {author} {\bibfnamefont {W.~G.}\ \bibnamefont {Lynch}}, \ and\ \bibinfo
  {author} {\bibfnamefont {M.~B.}\ \bibnamefont {Tsang}},\ }\href {\doibase
  10.1103/PhysRevC.68.051601} {\bibfield  {journal} {\bibinfo  {journal} {Phys.
  Rev. C}\ }\textbf {\bibinfo {volume} {68}},\ \bibinfo {pages} {051601}
  (\bibinfo {year} {2003})}\BibitemShut {NoStop}%
\bibitem [{\citenamefont {Trippa}\ \emph {et~al.}(2008)\citenamefont {Trippa},
  \citenamefont {Col\`o},\ and\ \citenamefont {Vigezzi}}]{Tri08}%
  \BibitemOpen
  \bibfield  {author} {\bibinfo {author} {\bibfnamefont {L.}~\bibnamefont
  {Trippa}}, \bibinfo {author} {\bibfnamefont {G.}~\bibnamefont {Col\`o}}, \
  and\ \bibinfo {author} {\bibfnamefont {E.}~\bibnamefont {Vigezzi}},\ }\href
  {\doibase 10.1103/PhysRevC.77.061304} {\bibfield  {journal} {\bibinfo
  {journal} {Phys. Rev. C}\ }\textbf {\bibinfo {volume} {77}},\ \bibinfo
  {pages} {061304} (\bibinfo {year} {2008})}\BibitemShut {NoStop}%
\bibitem [{\citenamefont {Furnstahl}(2002)}]{Fur02}%
  \BibitemOpen
  \bibfield  {author} {\bibinfo {author} {\bibfnamefont {R.}~\bibnamefont
  {Furnstahl}},\ }\href {\doibase
  https://doi.org/10.1016/S0375-9474(02)00867-9} {\bibfield  {journal}
  {\bibinfo  {journal} {Nucl. Phys. A}\ }\textbf {\bibinfo {volume} {706}},\
  \bibinfo {pages} {85 } (\bibinfo {year} {2002})}\BibitemShut {NoStop}%
\bibitem [{\citenamefont {Xie}\ and\ \citenamefont {Li}(2019)}]{Xie19}%
  \BibitemOpen
  \bibfield  {author} {\bibinfo {author} {\bibfnamefont {W.~J.}\ \bibnamefont
  {Xie}}\ and\ \bibinfo {author} {\bibfnamefont {B.~A.}\ \bibnamefont {Li}},\
  }\href {\doibase 10.3847/1538-4357/ab3f37} {\bibfield  {journal} {\bibinfo
  {journal} {Astrophys. J.}\ }\textbf {\bibinfo {volume} {883}},\ \bibinfo
  {pages} {174} (\bibinfo {year} {2019})}\BibitemShut {NoStop}%
\bibitem [{\citenamefont {Li}\ \emph {et~al.}(2008)\citenamefont {Li},
  \citenamefont {Chen},\ and\ \citenamefont {Ko}}]{BaLi08}%
  \BibitemOpen
  \bibfield  {author} {\bibinfo {author} {\bibfnamefont {B.~A.}\ \bibnamefont
  {Li}}, \bibinfo {author} {\bibfnamefont {L.~W.}\ \bibnamefont {Chen}}, \ and\
  \bibinfo {author} {\bibfnamefont {C.~M.}\ \bibnamefont {Ko}},\ }\href
  {\doibase https://doi.org/10.1016/j.physrep.2008.04.005} {\bibfield
  {journal} {\bibinfo  {journal} {Phys. Rep.}\ }\textbf {\bibinfo {volume}
  {464}},\ \bibinfo {pages} {113 } (\bibinfo {year} {2008})}\BibitemShut
  {NoStop}%
\bibitem [{\citenamefont {Horowitz}\ \emph {et~al.}(2014)\citenamefont
  {Horowitz}, \citenamefont {Brown}, \citenamefont {Kim}, \citenamefont
  {Lynch}, \citenamefont {Michaels}, \citenamefont {Ono}, \citenamefont
  {Piekarewicz}, \citenamefont {Tsang},\ and\ \citenamefont {Wolter}}]{Hor14}%
  \BibitemOpen
  \bibfield  {author} {\bibinfo {author} {\bibfnamefont {C.~J.}\ \bibnamefont
  {Horowitz}}, \bibinfo {author} {\bibfnamefont {E.~F.}\ \bibnamefont {Brown}},
  \bibinfo {author} {\bibfnamefont {Y.}~\bibnamefont {Kim}}, \bibinfo {author}
  {\bibfnamefont {W.~G.}\ \bibnamefont {Lynch}}, \bibinfo {author}
  {\bibfnamefont {R.}~\bibnamefont {Michaels}}, \bibinfo {author}
  {\bibfnamefont {A.}~\bibnamefont {Ono}}, \bibinfo {author} {\bibfnamefont
  {J.}~\bibnamefont {Piekarewicz}}, \bibinfo {author} {\bibfnamefont {M.~B.}\
  \bibnamefont {Tsang}}, \ and\ \bibinfo {author} {\bibfnamefont {H.~H.}\
  \bibnamefont {Wolter}},\ }\href {\doibase 10.1088/0954-3899/41/9/093001}
  {\bibfield  {journal} {\bibinfo  {journal} {J. Phys. G}\ }\textbf {\bibinfo
  {volume} {41}},\ \bibinfo {pages} {093001} (\bibinfo {year}
  {2014})}\BibitemShut {NoStop}%
\bibitem [{\citenamefont {Lattimer}(2014)}]{Latt14}%
  \BibitemOpen
  \bibfield  {author} {\bibinfo {author} {\bibfnamefont {J.~M.}\ \bibnamefont
  {Lattimer}},\ }\href {\doibase
  https://doi.org/10.1016/j.nuclphysa.2014.04.008} {\bibfield  {journal}
  {\bibinfo  {journal} {Nucl. Phys. A}\ }\textbf {\bibinfo {volume} {928}},\
  \bibinfo {pages} {276 } (\bibinfo {year} {2014})}\BibitemShut {NoStop}%
\bibitem [{\citenamefont {Garg}\ and\ \citenamefont {Colò}(2018)}]{Garg18}%
  \BibitemOpen
  \bibfield  {author} {\bibinfo {author} {\bibfnamefont {U.}~\bibnamefont
  {Garg}}\ and\ \bibinfo {author} {\bibfnamefont {G.}~\bibnamefont {Colò}},\
  }\href {\doibase https://doi.org/10.1016/j.ppnp.2018.03.001} {\bibfield
  {journal} {\bibinfo  {journal} {Prog. Part. Nucl. Phys.}\ }\textbf {\bibinfo
  {volume} {101}},\ \bibinfo {pages} {55 } (\bibinfo {year}
  {2018})}\BibitemShut {NoStop}%
\bibitem [{\citenamefont {Khoa}\ \emph
  {et~al.}(2007{\natexlab{b}})\citenamefont {Khoa}, \citenamefont {von
  Oertzen}, \citenamefont {Bohlen},\ and\ \citenamefont {Ohkubo}}]{Kho07r}%
  \BibitemOpen
  \bibfield  {author} {\bibinfo {author} {\bibfnamefont {D.~T.}\ \bibnamefont
  {Khoa}}, \bibinfo {author} {\bibfnamefont {W.}~\bibnamefont {von Oertzen}},
  \bibinfo {author} {\bibfnamefont {H.~G.}\ \bibnamefont {Bohlen}}, \ and\
  \bibinfo {author} {\bibfnamefont {S.}~\bibnamefont {Ohkubo}},\ }\href
  {\doibase 10.1088/0954-3899/34/3/r01} {\bibfield  {journal} {\bibinfo
  {journal} {J. Phys. G}\ }\textbf {\bibinfo {volume} {34}},\ \bibinfo {pages}
  {R111} (\bibinfo {year} {2007}{\natexlab{b}})}\BibitemShut {NoStop}%
\bibitem [{\citenamefont {Zhang}\ \emph {et~al.}(2018)\citenamefont {Zhang},
  \citenamefont {Li},\ and\ \citenamefont {Xu}}]{Zhang18}%
  \BibitemOpen
  \bibfield  {author} {\bibinfo {author} {\bibfnamefont {N.~B.}\ \bibnamefont
  {Zhang}}, \bibinfo {author} {\bibfnamefont {B.~A.}\ \bibnamefont {Li}}, \
  and\ \bibinfo {author} {\bibfnamefont {J.}~\bibnamefont {Xu}},\ }\href
  {\doibase 10.3847/1538-4357/aac027} {\bibfield  {journal} {\bibinfo
  {journal} {Astrophys. J.}\ }\textbf {\bibinfo {volume} {859}},\ \bibinfo
  {pages} {90} (\bibinfo {year} {2018})}\BibitemShut {NoStop}%
\bibitem [{\citenamefont {Pearson}\ \emph {et~al.}(2018)\citenamefont
  {Pearson}, \citenamefont {Chamel}, \citenamefont {Potekhin}, \citenamefont
  {Fantina}, \citenamefont {Ducoin}, \citenamefont {Dutta},\ and\ \citenamefont
  {Goriely}}]{Chamel18}%
  \BibitemOpen
  \bibfield  {author} {\bibinfo {author} {\bibfnamefont {J.~M.}\ \bibnamefont
  {Pearson}}, \bibinfo {author} {\bibfnamefont {N.}~\bibnamefont {Chamel}},
  \bibinfo {author} {\bibfnamefont {A.~Y.}\ \bibnamefont {Potekhin}}, \bibinfo
  {author} {\bibfnamefont {A.~F.}\ \bibnamefont {Fantina}}, \bibinfo {author}
  {\bibfnamefont {C.}~\bibnamefont {Ducoin}}, \bibinfo {author} {\bibfnamefont
  {A.~K.}\ \bibnamefont {Dutta}}, \ and\ \bibinfo {author} {\bibfnamefont
  {S.}~\bibnamefont {Goriely}},\ }\href {\doibase 10.1093/mnras/sty2413}
  {\bibfield  {journal} {\bibinfo  {journal} {Mon. Not. R. Astron. Soc.}\
  }\textbf {\bibinfo {volume} {481}},\ \bibinfo {pages} {2994} (\bibinfo {year}
  {2018})}\BibitemShut {NoStop}%
\bibitem [{\citenamefont {Mutafchieva}\ \emph {et~al.}(2019)\citenamefont
  {Mutafchieva}, \citenamefont {Chamel}, \citenamefont {Stoyanov},
  \citenamefont {Pearson},\ and\ \citenamefont {Mihailov}}]{Chamel19}%
  \BibitemOpen
  \bibfield  {author} {\bibinfo {author} {\bibfnamefont {Y.~D.}\ \bibnamefont
  {Mutafchieva}}, \bibinfo {author} {\bibfnamefont {N.}~\bibnamefont {Chamel}},
  \bibinfo {author} {\bibfnamefont {Z.~K.}\ \bibnamefont {Stoyanov}}, \bibinfo
  {author} {\bibfnamefont {J.~M.}\ \bibnamefont {Pearson}}, \ and\ \bibinfo
  {author} {\bibfnamefont {L.~M.}\ \bibnamefont {Mihailov}},\ }\href {\doibase
  10.1103/PhysRevC.99.055805} {\bibfield  {journal} {\bibinfo  {journal} {Phys.
  Rev. C}\ }\textbf {\bibinfo {volume} {99}},\ \bibinfo {pages} {055805}
  (\bibinfo {year} {2019})}\BibitemShut {NoStop}%
\bibitem [{\citenamefont {Douchin}\ and\ \citenamefont
  {Haensel}(2001)}]{Dou01}%
  \BibitemOpen
  \bibfield  {author} {\bibinfo {author} {\bibfnamefont {F.}~\bibnamefont
  {Douchin}}\ and\ \bibinfo {author} {\bibfnamefont {P.}~\bibnamefont
  {Haensel}},\ }\href {\doibase 10.1051/0004-6361:20011402} {\bibfield
  {journal} {\bibinfo  {journal} {Astron. Astrophys.}\ }\textbf {\bibinfo
  {volume} {380}},\ \bibinfo {pages} {151} (\bibinfo {year}
  {2001})}\BibitemShut {NoStop}%
\bibitem [{\citenamefont {Demorest}\ \emph {et~al.}(2010)\citenamefont
  {Demorest}, \citenamefont {Pennucci}, \citenamefont {Ransom}, \citenamefont
  {Roberts},\ and\ \citenamefont {Hessels}}]{Demorest}%
  \BibitemOpen
  \bibfield  {author} {\bibinfo {author} {\bibfnamefont {P.~B.}\ \bibnamefont
  {Demorest}}, \bibinfo {author} {\bibfnamefont {T.}~\bibnamefont {Pennucci}},
  \bibinfo {author} {\bibfnamefont {S.~M.}\ \bibnamefont {Ransom}}, \bibinfo
  {author} {\bibfnamefont {M.~S.~E.}\ \bibnamefont {Roberts}}, \ and\ \bibinfo
  {author} {\bibfnamefont {J.~W.~T.}\ \bibnamefont {Hessels}},\ }\href
  {\doibase 10.1038/nature09466} {\bibfield  {journal} {\bibinfo  {journal}
  {Nature}\ }\textbf {\bibinfo {volume} {467}},\ \bibinfo {pages} {1081}
  (\bibinfo {year} {2010})}\BibitemShut {NoStop}%
\bibitem [{\citenamefont {Arzoumanian~\textit{et al.}}(2018)}]{Arzou18}%
  \BibitemOpen
  \bibfield  {author} {\bibinfo {author} {\bibfnamefont {Z.}~\bibnamefont
  {Arzoumanian~\textit{et al.}}},\ }\href {\doibase 10.3847/1538-4365/aab5b0}
  {\bibfield  {journal} {\bibinfo  {journal} {Astrophys. J. Suppl. Ser.}\
  }\textbf {\bibinfo {volume} {235}},\ \bibinfo {pages} {37} (\bibinfo {year}
  {2018})}\BibitemShut {NoStop}%
\bibitem [{\citenamefont {Antoniadis~\textit{et al.}}(2013)}]{Sci340}%
  \BibitemOpen
  \bibfield  {author} {\bibinfo {author} {\bibfnamefont {J.}~\bibnamefont
  {Antoniadis~\textit{et al.}}},\ }\href
  {https://science.sciencemag.org/content/340/6131/1233232} {\bibfield
  {journal} {\bibinfo  {journal} {Science}\ }\textbf {\bibinfo {volume} {340}}
  (\bibinfo {year} {2013})}\BibitemShut {NoStop}%
\bibitem [{\citenamefont {Cromartie}\ \emph {et~al.}(2020)\citenamefont
  {Cromartie}, \citenamefont {Fonseca},\ and\ \citenamefont {Ransom~\textit{et
  al.}}}]{Cro19}%
  \BibitemOpen
  \bibfield  {author} {\bibinfo {author} {\bibfnamefont {H.~T.}\ \bibnamefont
  {Cromartie}}, \bibinfo {author} {\bibfnamefont {E.}~\bibnamefont {Fonseca}},
  \ and\ \bibinfo {author} {\bibfnamefont {S.~M.}\ \bibnamefont
  {Ransom~\textit{et al.}}},\ }\href {\doibase 10.1038/s41550-019-0880-2}
  {\bibfield  {journal} {\bibinfo  {journal} {Nat. Astron.}\ }\textbf {\bibinfo
  {volume} {4}},\ \bibinfo {pages} {72} (\bibinfo {year} {2020})}\BibitemShut
  {NoStop}%
\bibitem [{\citenamefont {Tsang}\ \emph {et~al.}(2019)\citenamefont {Tsang},
  \citenamefont {Tsang}, \citenamefont {Danielewicz}, \citenamefont
  {Fattoyev},\ and\ \citenamefont {Lynch}}]{Tsang19}%
  \BibitemOpen
  \bibfield  {author} {\bibinfo {author} {\bibfnamefont {C.~Y.}\ \bibnamefont
  {Tsang}}, \bibinfo {author} {\bibfnamefont {M.~B.}\ \bibnamefont {Tsang}},
  \bibinfo {author} {\bibfnamefont {P.}~\bibnamefont {Danielewicz}}, \bibinfo
  {author} {\bibfnamefont {F.~J.}\ \bibnamefont {Fattoyev}}, \ and\ \bibinfo
  {author} {\bibfnamefont {W.~G.}\ \bibnamefont {Lynch}},\ }\href {\doibase
  https://doi.org/10.1016/j.physletb.2019.05.055} {\bibfield  {journal}
  {\bibinfo  {journal} {Phys. Lett. B}\ }\textbf {\bibinfo {volume} {796}},\
  \bibinfo {pages} {1 } (\bibinfo {year} {2019})}\BibitemShut {NoStop}%
\end{thebibliography}%
\end{document}